\documentclass{aastex631}

\usepackage{threeparttable}

\newcommand{\kms}{\,km\,s$^{-1}$\ }
\newcommand{\HII}{\textsc{Hii}\ }

\begin{document}

\title{A New Cloud-Cloud Collision Source N68 toward the G35 Molecular Cloud Complex}

\author[0009-0002-8449-1734]{En Chen}
\affiliation{Center for Astrophysics, Guangzhou University, Guangzhou 510006, P. R. China}

\author[0000-0002-5435-925X]{Xi Chen}
\affiliation{Center for Astrophysics, Guangzhou University, Guangzhou 510006, P. R. China}

\correspondingauthor{Xi Chen}
\email{chenxi@gzhu.edu.cn}



\begin{abstract}

Bubble N68 in the G35 complex shows clear cloud-cloud collision (CCC) signatures. 
Its semi-ring-like morphology harbors many significant massive star formation tracers: 6 \HII regions, 4 6.7 GHz masers, 5 Midcourse Space Experiment sources, 9 radio peaks, and nearly 10 O/B-type stars. 
We also identified 163 young stellar objects (45 Class I, 5 Flat, 113 Class II), indicating active star formation toward N68. 
Our molecular study with CO reveals two distinct molecular clouds (N68a: 47-56 \kms; N68b: 56-64 \kms), with broad bridge features and complementary distributions at their borders, indicating an ongoing CCC. 
Star formation in N68 is collectively driven by collect-and-collapse (CC), radiation-driven implosion (RDI), and CCC mechanisms. 
However, compared with the CC and RDI mechanisms, the CCC mechanism does not enhance the star formation efficiency; instead, it tends to trigger the formation of massive stars.
N68, along with bubbles N65 and N61, constructs a $\sim100$ pc scale CCC system in the G35 complex. 

\end{abstract}



\section{Introduction} \label{sec:intro}

Massive stars ($\geq8$ M$_{\odot}$) are prominent in the Galaxy because of their extreme density, high luminosity, and violent interaction with its surrounding interstellar medium (ISM). 
Despite the importance of massive star formation to the evolution of galaxies, there is currently no certain theory to explain its formation mechanism. 
Observationally, there are many signatures indicating the existence of massive star formation, for example, infall/outflow of dense cores \citep{Klaassen2007}, masers \citep{Urquhart2013}, compact radio sources \citep{Masque2017} and (Ultra-Compact; UC) \HII regions \citep{Churchwell2002}, etc. 
These indicators show massive star formation activity on molecular clouds. 
Among them, infall/outflow and masers indicate the early stages of massive star formation, whereas radio sources and (UC) \HII regions show its evolved stages. 
However, the massive star formation process remains less understood than that of their low-mass counterparts, primarily because of their rarity, short lifetime, deep embedded, and significant feedback. 

Recent observations seem to suggest that triggered star formation, such as the \HII region expansion and the cloud-cloud collision (CCC) process, may be important mechanisms for massive star formation. 
The \HII region, as a trigger, usually performs two processes to induce star formation due to its dynamics and ionization feedback: collect-and-collapse (CC; \citealt{Elmegreen1977}), and radiation-driven implosion (RDI; \citealt{Sandford1982}). 
In the CC process, due to the expansion of the \HII region, its ionization front and shock front will compress its surrounding gas and collect to form a dense shell, which further fragments and collapses to form stars \citep{Whitworth1994, Deharveng2005, Deharveng2010}. 
\cite{Dale2007a} repeated the simulation based on the \cite{Whitworth1994} model, and found that the shell driven by the \HII region fragments to form numerous self-gravitating objects, supporting the CC model. 
RDI refers to the ionization shock front that sweeps across the pre-existing density structures, forming stars characterized by cometary globules or optically bright rims \citep{Sugitani1991, Bisbas2011}. 
On the other hand, the CCC process, as an external trigger, can rapidly accumulate mass into a small volume, resulting in the formation of ultra-dense massive cores that continue to form massive stars by self-gravity. 
The simulation conducted by \cite{Inoue2013} indicates that the strong shock waves generated by CCC can trigger the formation of massive, gravitationally unstable molecular cloud cores. 
This conclusion is also supported by many observations. 
\cite{Fukui2021} suggested that almost all massive star-forming regions (MSFRs) in the Milky Way, such as M42, NGC 3603, NGC 6334, RCW 38, etc., are triggered by CCC processes. 
In addition, the CCC process is also regarded as a multi-functional mechanism, which can trigger the formation of low-/intermediate-mass stars \citep{Gong2017}, individual O-type stars \citep{Torii2015, Torii2017}, as well as super star clusters (SSCs) or starbursts \citep{Fukui2014, Fukui2016} containing tens of O-type stars.

\begin{figure}
	\includegraphics[width=1\textwidth]{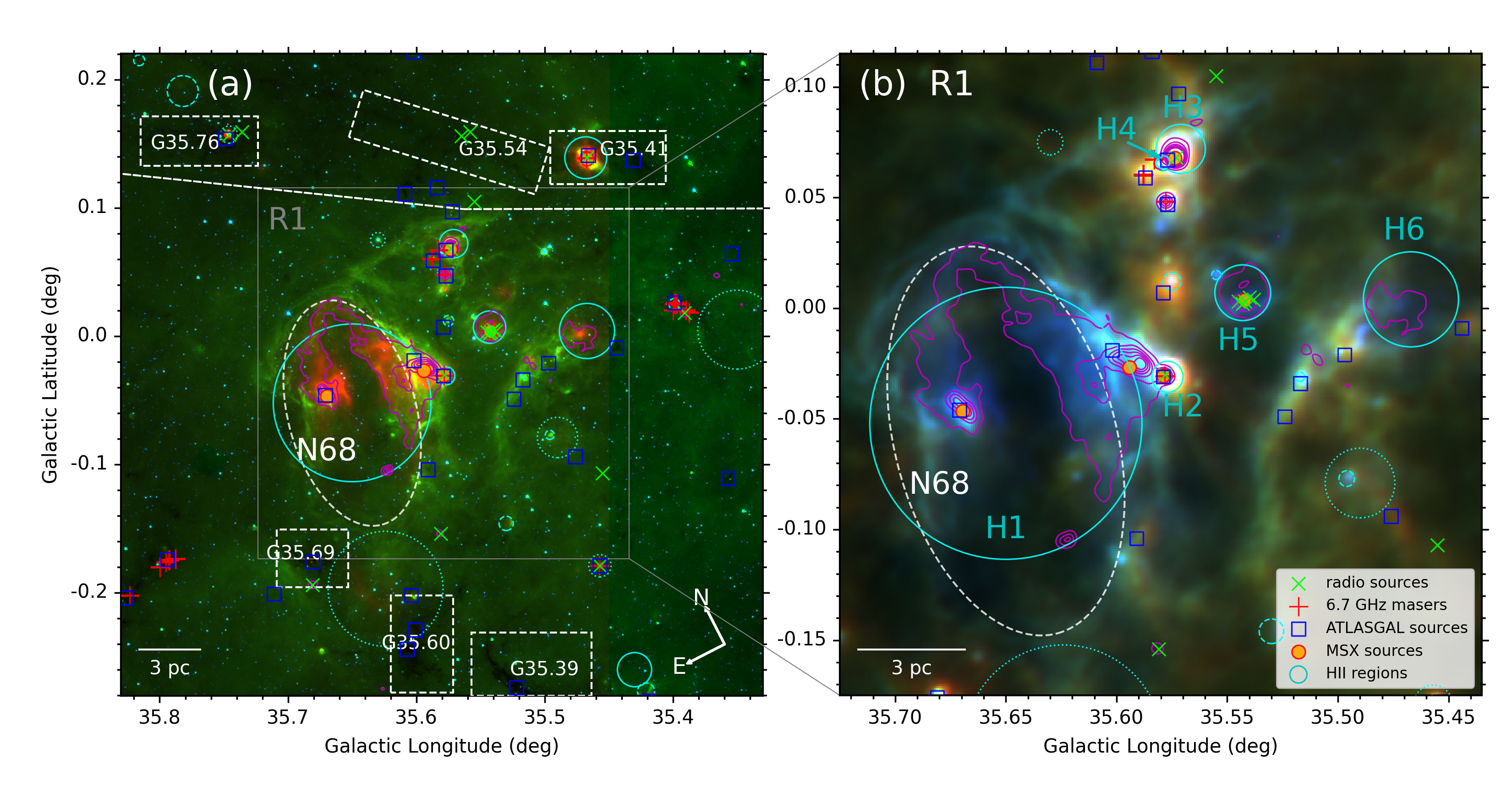}
    \caption{Overview of the G35 complex. Panel (a) shows a $Spitzer$ RGB-composite image (24, 8, and 3.6 $\mu$m) of the region surrounding MIR bubble N68. Panel (b) zooms in on region R1 in a $Herschel$ RGB-composite (350, 160, and 70 $\mu$m). Overlaid on both panels are tracers of massive star formation: MAGPIS radio sources (green crosses), 6.7 GHz methanol masers (red pluses), ATLASGAL 870 $\mu$m peaks (blue squares), $MSX$ MYSO candidates (orange dots), and WISE \HII regions (cyan circles). The 20 cm radio continuum (magenta contours; levels start from 0.0008 to 0.0088 in steps of 0.0016 Jy beam$^{-1}$) traces the inner rim of bubble N68, outlined by a white dashed ellipse. White dashed rectangles mark dark clouds identified in \cite{Shen2024}.}
    \label{fig:rgb}
\end{figure}

The mid-infrared (MIR) bubble N68 is selected as the target region that was detected and cataloged by \cite{Churchwell2006}. 
As shown in Figure \ref{fig:rgb}, the target region, centered at $(l, b)= (35\fdg578, -0\fdg03)$, with a box size of $0\fdg3$, is part of the giant molecular cloud (GMC) G35 complex. 
\cite{Shen2024} proposed that the complex contains five individual clouds with different systemic velocities (designated as Clouds $1-5$), many filamentary structures, and infrared dark clouds (IRDCs; i.e., G35.39, G35.60, G35.69, G35.41, G35.54, and G35.46) at different evolutionary stages. 
Our target region (R1 in Figure \ref{fig:rgb} (b)) corresponds to their Cloud 4, with a velocity range of 40 to 65 \kms. 
This cloud is very bright in the IR bands, containing a large infrared bubble N68 and several small \HII regions (marked with cyan circles in Figure \ref{fig:rgb};  \citealt{Anderson2014}). 
Six Red $MSX$ Sources (RMS; \citealt{Urquhart2008, Lumsden2013}, marked by orange dots) are detected in the target region, which were considered massive young stellar objects (MYSOs). 
They are mainly distributed in the periphery of N68 and the center of the small \HII regions (H2, H3 and H5), indicating massive star formation there. 
The elliptic bubble structure of N68 might be caused by the combined effect of stellar feedback from multiple O/B-type stars. 
The extended and asymmetric radio continuum emissions (MAGPIS 20 cm; \citealt{Helfand2006}) provide a clue to this. 
Four methanol masers at 6.7 GHz suggest the early stage of high-mass star formation, which were observed by the Global View on Star Formation (GLOSTAR) survey with high-resolution Very Large Array (VLA) observations \citep{Brunthaler2021, Nguyen2022}. 
Our target region contains a total of 13 \HII regions, among which 6 are classified as ``known" (marked by H1$-$H6), 4 are classified as ``radio quiet", and 3 are classified as ``candidate" in the WISE \HII region catalog of \cite{Anderson2014}. 
All of the above indicators suggest that the entire target region is a highly active MSFR, which encompasses the formation of massive stars at various evolutionary phases, from the initial stage with masers to the evolved bubbles or \HII regions.

In the literature, there is no robust distance measurement for bubble N68 or its associated counterparts. 
The simplest and most common way to estimate the distance to a molecular cloud is to convert its radial velocity into kinematic distance based on the Galactic rotation model. 
However, this method will lead to the problem of kinematic distance ambiguity (KDA) toward the direction of inner Galaxy, for one velocity corresponds to two possible distances (a ``near" and a ``far" distance). 
For example, using the Galaxy parameter model A5 in \cite{Reid2014}, the typical velocity 53.8 \kms of N68 corresponds to a ``near" distance of 3.2 kpc and a ``far" distance of 10.2 kpc, respectively. 
\cite{Anderson2009} obtained the ``far" distance for N68 based on comparing the velocity of the ionized gas with the maximum velocity of \textsc{Hi} absorption. 
\cite{Zhang2013} involved this distance of 10.6 kpc in their study and derived the size of bubble N68 of 34 pc $\times$17 pc. 
However, the typical radius of an infrared bubble is less than 15 pc \citep{Churchwell2006}, therefore, this ``far" distance of N68 may not be appropriate for deriving physical properties. 
\cite{Shen2024} employed the latest distance calculation tool, the ``Parallax-Based Distance Calculator V2" \citep{Reid2016, Reid2019}, to derive the kinematic distance of their Cloud 1, Cloud 3 and Cloud 4 (together associated with the same complex) of 2.5 kpc. 
Taking into account the distance of the well-studied IRDC G35.39 (2.9 kpc, lied in Cloud 1), they finally adopted the distance of Cloud 4 (N68) as 2.9 kpc, consistent with the ``near" side distance. 
Furthermore, in this paper, we will argue that bubbles N68, N65 (3.5 kpc; \citealt{Chen2025}) and N61 (3.7 kpc; \citealt{Xu2016}) may be associated at a scale of 100 pc, hence, we tend to consider that N68 is also at a ``near" distance. 
In conclusion, N68 is more likely to be located within the ``near" distance between 2.9 and 3.5 kpc. 
Considering the association relationship between N68 and N65 within 100 pc, we finally set the distance of N68 at 3.4 kpc for consistency. 
Based on this, we derive the size of N68 as 11 pc $\times$ 6 pc, which is a typical size of MIR bubbles. 

The paper consists of five sections. 
Section \ref{sec:observations} describes the target and the CO observations and the archival data we used. 
Section \ref{sec:results} presents the structure and kinematics of molecular
clouds and star formation activities in the study region. 
Section \ref{sec:discussion} discusses the evidence for CCC and its effects on
triggering star formation. 
And finally, Section \ref{sec:conclusion} gives the conclusions of this work.

\section{Observations} \label{sec:observations}

\subsection{MWISP CO observations} \label{subsec:MWISP}

The observations of CO toward the MIR bubble N68 are part of the Milky Way Imaging Scroll Painting (MWISP\footnote{\url{http://www.radioast.nsdc.cn/mwisp.php/}}) project (see \citealt{Su2019}) conducted by the PMO-13.7m single-dish millimeter telescope at Delingha in China. 
The antenna employed the $3\times3-$beam Superconducting Spectroscopic Array Receiver (SSAR) system as its front end \citep{Shan2012} to obtain three molecular lines of CO, $^{12}$CO, $^{13}$CO, and C$^{18}$O ($J=1-0$) simultaneously under the on-the-fly (OTF) mode. 
A total of 18 Fast Fourier Transform (FFT) spectrometers are used as its back end that provide a 1 GHz bandwidth with 16384 channels, producing a frequency interval of 61 kHz and a velocity coverage of $\sim$ 2600 km s$^{-1}$. 
The channel separation and the RMS noise level in a channel width are 0.158 km s$^{-1}$ and 0.48 K for $^{12}$CO data, and 0.166 km s$^{-1}$ and 0.3 K for $^{13}$CO and C$^{18}$O data. 
The telescope has a beam size of 55$^{\prime\prime}$ at 110 GHz ($^{13}$CO and C$^{18}$O) and 52$^{\prime\prime}$ at 115 GHz ($^{12}$CO), with a pointing accuracy of $\sim 5^{\prime\prime}$. 
The final data product was converted into 3D FITS data cubes of each cell ($30^{\prime}\times30^{\prime}$) with a grid spacing of $30^{\prime\prime}$. 

The Nobeyama 45m telescope has provided similar CO data (FUGIN, \citealt{Umemoto2017}), with an angular resolution reaching $20^{\prime\prime}$, but it has a lower velocity resolution of 1.3 \kms. 
Since the aim of this study is to resolve the velocity bridge feature (1-3 \kms) between molecular clouds, the MWISP data with higher velocity resolution can better meet our requirements. 
In addition, the angular resolution of the MWISP data is also sufficient to resolve the collision interface (3-5$^{\prime}$) on N68. 
Therefore, the MWISP data was adopted here instead of the FUGIN data.

In this paper, the data cubes are clipped into a box field of $18^{\prime}\times18^{\prime}$ centered at ($l, b$) $\sim$ ($35\fdg578$, $-0\fdg03$) and a velocity range of $0\sim140$ km s$^{-1}$. 
Throughout this paper, the Galactic coordinate system is utilized and the equinox is J2000.0, and the velocities are all given with respect to the local standard of rest (LSR).

\subsection{Archival data} \label{subsec:archive}

Table \ref{tab:surveys} presents a summary of the archival data used in this paper, with their wavelengths spinning from near-infrared (NIR) to radio emission. 
The NIR J-, H- and K-band photometric data from the UKIRT Infrared Deep Sky Survey (UKIDSS) DR10PLUS data release \citep{Lawrence2007} and the MIR photometric data from the $Spitzer$ space telescope \citep{Fazio2004} are combined to select the disk-bearing young stellar object (YSO) candidates in the study region. 
The MIR images at IRAC 3.6, 4.5, 5.8 and 8.0 $\mu$m from GLIMPSE and MIPS 24 $\mu$m from MIPSGAL are used to trace the emission from warm dust. 
And the $Herschel$ images at 70, 160, 250, 350, and 500 $\mu$m from the $Herschel$ infrared Galactic Plane Survey (Hi-GAL) are adopted to trace the emission from cold dust. 
The submillimeter continuum emission at 870 $\mu$m from the APEX Telescope Large Area Survey of the Galaxy (ATLASGAL, \citealt{Schuller2009}) is used to study the dust clumps. 
The radio continuum emissions that trace the ionized gas are obtained from the radio interferometer surveys of the Galactic Plane, namely CORNISH (5 GHz; \citealt{Hoare2012}), MAGPIS (1.4 GHz; \citealt{Helfand2006}) and NVSS (1.4 GHz; \citealt{Condon1998}). 
The higher resolution CORNISH (1.5$^{\prime\prime}$) and MAGPIS (6$^{\prime\prime}$) focus on capturing compact sources, while the lower resolution NVSS (46$^{\prime\prime}$) perform exceptionally well in capturing extended emissions. 
In addition, the 6.7 GHz methanol masers that trace the early stage of massive star formation were identified using the online tool $MaserDB$\footnote{\url{https://maserdb.net}}, provided by \cite{Ladeyschikov2019}.

\setlength{\tabcolsep}{3pt}
\begin{table*}
    \tiny
    \centering
    \caption{Catalog of various surveys utilized in this study, covering a wide range of wavelengths from near-infrared to radio.} \label{tab:surveys}
    \begin{threeparttable}
    \begin{tabular}{cccc}
        \hline
        \hline
        Survey & Wavelengths/Bands & Resolution & Reference \\
        \hline
        The UKIRT Infrared Deep Sky Survey (UKIDSS)\tnote{a} & 1.2, 1.6, 2.2 $\mu$m & 0.8$^{\prime\prime}$, 0.8$^{\prime\prime}$, 0.8$^{\prime\prime}$ & \cite{Lawrence2007} \\
        Galactic Legacy Infrared Mid-plane Survey Extraordinaire (GLIMPSE)\tnote{b} & 3.6, 4.5, 5.8, 8.0 $\mu$m & $\sim2^{\prime\prime}$, $\sim2^{\prime\prime}$, $\sim2^{\prime\prime}$, $\sim2^{\prime\prime}$ & \cite{Benjamin2003} \\
        A 24 and 70 Micron Survey of the Inner Galactic Disk with MIPS (MIPSGAL)\tnote{c} & 24, 70 $\mu$m & 6$^{\prime\prime}$, 18$^{\prime\prime}$ & \cite{Carey2009} \\
        The $Herschel$ Infrared Galactic Plane Survey (Hi-GAL)\tnote{d} & 70, 160, 250, 350, 500 $\mu$m & 5.8$^{\prime\prime}$, 12$^{\prime\prime}$, 18$^{\prime\prime}$, 25$^{\prime\prime}$, 37$^{\prime\prime}$ & \cite{Molinari2010} \\
        The Milky Way Imaging Scroll Painting (MWISP)\tnote{e} & 260, 272, 273 $\mu$m & 52$^{\prime\prime}$, 55$^{\prime\prime}$, 55$^{\prime\prime}$ & \cite{Su2019} \\
        The APEX Telescope Large Area Survey of the Galaxy (ATLASGAL)\tnote{f} & 870 $\mu$m & 36$^{\prime\prime}$ & \cite{Schuller2009} \\
        Co-Ordinated Radio `N' Infrared Survey for High-mass star formation (CORNISH)\tnote{g} & 6 cm & 1.5$^{\prime\prime}$ & \cite{Hoare2012} \\
        Multi-Array Galactic Plane Imaging Survey (MAGPIS)\tnote{h} & 20 cm & 6$^{\prime\prime}$ & \cite{Helfand2006} \\
        NRAO VLA Sky Survey (NVSS)\tnote{i} & 21 cm & 46$^{\prime\prime}$ & \cite{Condon1998} \\
        \hline
    \end{tabular}
    \noindent\parbox{\textwidth}{
    \tiny
    \textsc{Note}---$^{\rm a}$\url{http://ukidss.org}; 
    $^{\rm b}$\url{https://irsa.ipac.caltech.edu/data/SPITZER/GLIMPSE/overview.html}; 
    $^{\rm c}$\url{https://irsa.ipac.caltech.edu/data/SPITZER/MIPSGAL/overview.html}; 
    $^{\rm d}$\url{https://archives.esac.esa.int/hsa/whsa/}; 
    $^{\rm e}$\url{http://www.radioast.cn}; 
    $^{\rm f}$\url{https://www3.mpifr-bonn.mpg.de/div/atlasgal/}; 
    $^{\rm g}$\url{https://cornish.leeds.ac.uk/public/index.php}; 
    $^{\rm h}$\url{https://third.ucllnl.org/gps/index.html}; 
    $^{\rm i}$\url{https://www.cv.nrao.edu/nvss/anonftp.shtml}. 
    }
    \end{threeparttable}
\end{table*}

\section{Results} \label{sec:results}

\subsection{Ionized Emission and \HII Regions} \label{subsec:ionized}

\begin{figure}
    \centering
    \includegraphics[width=1\textwidth]{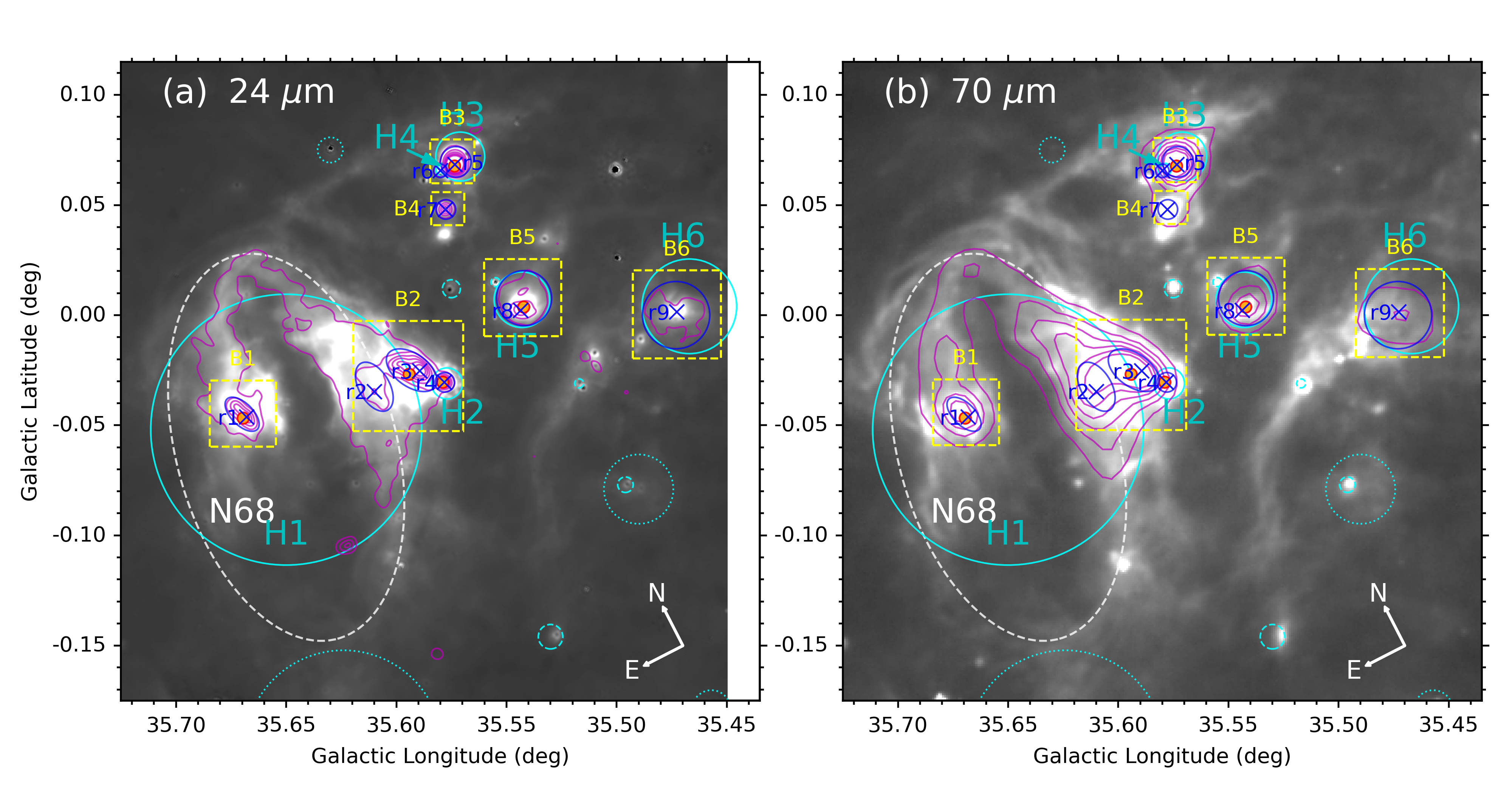}
    \caption{(a) $Spitzer$ 24 $\mu$m and (b) $Herschel$ 70 $\mu$m images with overlaid MAGPIS 20 cm contours (from 0.0008 to 0.0088 in steps of 0.0016 Jy beam$^{-1}$) and NVSS 21 cm contours (from 0.02 to 0.22 in steps of 0.04 Jy beam$^{-1}$) in magenta, respectively. The blue crosses and ellipses represent the peaks (markers from r1 to r9) and extents of the local clumps of the 20 cm radio continuum emission, respectively. These subregions are highlighted by yellow boxes with markers from B1 to B6. The konwn \HII regions are fitted by cyan circles with markers from H1 to H6. The other symbols are drawn as the same manners as in Figure \ref{fig:rgb}. }
    \label{fig:overlay}
\end{figure}

The multi-band data from infrared to radio reveal multiple signatures of active star formation in the N68 region.  
Figure \ref{fig:overlay}(a) and \ref{fig:overlay}(b) show the compact \HII regions in the region with MAGPIS (overlay on 24 $\mu$m emission) and NVSS (overlay on 70 $\mu$m emission), respectively. 
Both the higher resolution (6$^{\prime\prime}$) MAGPIS and the lower resolution (46$^{\prime\prime}$) NVSS data show six distinct subregions with strong radio continuum emissions, labeled as region B1$-$B6 (yellow boxes), indicating that these regions may embed exciting O/B-type stars. 
The fact that all these regions (except B4) are associated with known \HII regions (H1$-$H6; cyan circles) also proves this.  
The higher resolution MAGPIS radio continuum emission shows multiple local radio peaks, marked with r1$-$r9 (green crosses), which are located at the center of \HII regions, indicating the possible position of the exciting O/B stars. 
Notably, the radio continuum outlines the northern edge of the N68 bubble, presenting a ``n" or semi-ring-like shape structure, with B1 and B2 located in the eastern and western outer shells of N68, respectively. 
This asymmetry morphology of the bubble suggests that it is possibly caused by the combined effects of multiple exciting O/B stars. 
In addition, bubble N68 has also been identified as an evolved \HII region (i.e., H1), containing three radio peaks (r1, r2 and r3), indicating the possibility of the existence of multiple exciting stars. 
Interestingly, r1 (within region B1) is located at the eastern edge of the bubble, slightly offset from the center, with its ``head" pointing towards the bright emission region B2 in the western edge, presenting a typical bright rim cloud (BRC)-like structure. 
Such BRC structure is commonly interpreted as signatures of radiation-driven implosion (RDI) process, where an ionization front from a nearby source compresses a pre-existing molecular core. 
Further analysis and discussion of the BRC and associated star formation mechanisms are presented in Sections \ref{subsec:structure} and \ref{subsec:SF}.

Figure \ref{fig:subregions} presents detailed multi-wavelength views of the radio continuum subregions B1–B6, shown in 4.5, 8.0, 24, and 70 $\mu$m emission. 
All subregions display significant indicators of ongoing massive star formation. 
Specifically, radio peaks r1, r3, r4, r5, and r8 are spatially coincident with MYSO candidates identified as $MSX$ sources, while r4, r6, and r7 are associated with 6.7 GHz methanol masers—both clear tracers of high-mass star formation. 
Although r2 and r9 lack direct MYSO or maser detections, they exhibit strong 24 and 70 $\mu$m emission, characteristic of dust heated by embedded star-forming activity. 
In terms of ionized gas, r4, r5, r6, r8 and r9 correspond to known (ultra-) compact \HII regions (H2, H3, H4, H5 and H6, respectively). 
Radio peaks r1, r2, and r3 are located along the outer shell of the bubble and are linked to the evolved \HII region H1. 
Although r7 has not been formally cataloged as an \HII region, its association with a dense core and a 6.7 GHz maser suggests it may be a young UC \HII region in an early formation stage. 
As a result, the radio peaks r1–r9 are confirmed to be associated with \HII regions. 
Consequently, we can derive their physical properties following the methodology described in \cite{Chen2024} and \cite{Chen2025} based on the measured radius of each \HII region and the integrated radio continuum flux density.

In the methodology, we first calculate the hydrogen-ionizing photon rate $\dot{N}_{LyC}$ from the integrated flux density ($S_{\nu}$) of radio continuum emission as follows \citep{Condon1992}, 
\begin{equation}
    \dot{N}_{LyC} \approx 7.54\times10^{46}\ \Big(\frac{T_e}{10^4\ \rm K}\Big)^{-0.45}\Big(\frac{\nu}{\rm GHz}\Big)^{0.1}\Big(\frac{S_{\nu}}{\rm Jy}\Big)\Big(\frac{d}{\rm kpc}\Big)^2\ \rm s^{-1}, 
\end{equation}
where $T_e$ is the electron temperature (typical value of $10^4$ K), $\nu$ is the central frequency of the radio continuum emission (1.4 GHz for MAGPIS), and $d$ ($=3.4$ kpc) is the distance of the \textsc{Hii} region. 
Compared with Table 1 of \cite{Smith2002}, the spectral types of the exciting stars of these \HII regions can be determined, and the results are listed in Table \ref{tab:bubbles}.

For an evolved \textsc{Hii} region, we estimate its age using a dynamical age from \cite{Spitzer1978}, assuming spherical expansion, 
\begin{equation}
    t_{dyn} \approx 0.057\ \Big(\frac{R_s}{\rm pc}\Big)\Big(\frac{v_i}{\rm 10\ km\ s^{-1}}\Big)^{-1}\left(\Big(\frac{R}{R_s}\Big)^{7/4}-1\right)\ \rm Myr, 
\end{equation}
where $v_i$ is the sound speed of the ionized gas, typically $v_i\approx10$ km s$^{-1}$, $R$ is the radius of the evolved \textsc{Hii} region, and $R_s$ is the radius of the Str$\ddot{\rm o}$mgren sphere, which can be calculated with the formula \citep{Stromgren1939, Dyson1980}, 
\begin{equation}
    R_s = \Big(\frac{3\ \dot{N}_{LyC}}{4\pi\beta_*n_0^2}\Big)^{1/3} \approx 0.74\ \dot{N}_{49}^{1/3}\ n_3^{-2/3}\ \rm pc, 
\end{equation}
where $\beta_*\sim2\times10^{-13}$ cm$^{3}$ s$^{-1}$ is the recombination coefficient for atomic hydrogen. $n_3\equiv [n_0/10^3\ \rm cm^{-3}]$ is the initial number density of gas in a unit of 10$^{3}$ cm$^{-3}$, and $\dot{N}_{49}\equiv [\dot{N}_{LyC}/10^{49}\ \rm s^{-1}]$ is the hydrogen-ionizing photon rate in a unit of $10^{49}\ \rm s^{-1}$. 
Throughout this calculation, the initial number density of gas in a \HII region is assumed to be $5\times10^3$ cm$^{-3}$.

However, for a UC \HII region, since it is embedded within the dense core of its parent molecular cloud, no robust radius can be determined. 
Therefore, we estimate its age by assuming its radius as the Str$\ddot{\rm o}$mgren radius, according to \cite{Whitworth1994}, 
\begin{equation}
    t_s\approx0.042\ \dot{N}_{49}^{1/3}\ n_3^{-2/3}\ \rm Myr, 
\end{equation}
Based on their compact morphology in both radio continuum (tracing the inner ionized boundary) and 8 $\mu$m emission (tracing the outer photodissociation region), we classify the \HII regions r4, r6, and r7 as UC \HII regions. 
Using the method described above, we estimate their dynamical ages to be approximately 3, 2 and 2 kyr, respectively. 
Notably, all three regions share consistent physical characteristics: they are all embedded in a dense core and with a 6.7 GHz maser, indicating that they are in the early evolutionary stage of protostars with hot cores. 
Furthermore, r6 is spatially located on the shell structure of the adjacent r5, suggesting a potential triggering relationship between these two sources. 
More discussion will be presented in Section \ref{subsec:SF}. 

Based on the calculations presented above, we conclude that the N68 region hosts a significant population of O/B-type stars. 
As summarized in Table \ref{tab:bubbles}, bubble N68 originated from the expansion of the large \HII region H1, which is ionized by either an O8.5V star or multiple O9.5V stars. 
Although no central massive star is observed within the bubble, studies such as \cite{Hanaoka2019} indicate that the massive stars responsible for such structures are not necessarily located at their geometric centers.  
This evolved \HII region, with an estimated age of 2.9 Myr, appears to have triggered subsequent massive star formation along its periphery, giving rise to three B-type stars (r1, r2, and r4) and one O9.5V star (r3).  
An additional O9.5V star is identified at r5, which appears to have triggered the formation of a B1.5V star (r6) in its surrounding shell. 
In contrast, stars r7, r8, and r9 are relatively isolated B-type stars, and their formation may not have been directly influenced by external triggering mechanisms.

\begin{figure}
    \centering
    \includegraphics[width=0.78\textwidth]{show_subregions.png}
    \caption{Images of zoomed-in subregions B1 to B6 (as shown in Figure \ref{fig:overlay}) in different infrared bands (4.5, 8.0, 24 and 70 $\mu$m, respectively). The magenta contours indicate the MAGPIS 20 cm radio continuum emission, and the green crosses and ellipses just fit the peaks and extents of their local clumpy emission. The blue contours show the ATLASGAL 870 $\mu$m emission with their peaks marked by blue squares. The cyan circles indicate the six known \HII regions. The red pluses and orange dots indicate the 6.7 GHz masers and MYSOs, respectively. }
    \label{fig:subregions}
\end{figure}

\begin{deluxetable*}{cccccccccccc}
\tablecaption{Physical properties of the radio continuum emission subregions in N68. \label{tab:bubbles}} 
\tablehead{
\colhead{Region} & \colhead{\HII} & \colhead{$l$} & \colhead{$b$} & \colhead{R} & \colhead{S$_{\nu}$} & \colhead{log$(\dot{\rm N}_{LyC})$} & \colhead{Sp} & \colhead{R$_s$} & \colhead{t$_{dyn}$} & \colhead{R$_{frag}$} & \colhead{t$_{frag}$} \\
\cline{1-12}
\colhead{} & \colhead{ID} & \colhead{($^{\circ}$)} & \colhead{($^{\circ}$)} & \colhead{(pc $|$ $\prime\prime$)} & \colhead{(mJy)} & \colhead{(s$^{-1}$)} & \colhead{type} & \colhead{(pc)} & \colhead{(Myr)} & \colhead{(pc)} & \colhead{(Myr)} 
}
\colnumbers
\startdata 
         /           & H1   & 35.6500  &   -0.0600   & 4.09 (241)   & 2036$\pm$809 & 48.26 & O8.5V & 0.144 & 2.90 & 2.51 & 1.30 \\
  \cline{1-12}
  B1 \\
  \cline{1-1}
  \hspace{-3em} r1  & H1-a   & 35.6681 & -0.0461 & 0.42 (25) & 190$\pm$18  & 47.23 & B0V   & 0.065 & 0.10 & 2.02 & 1.52 \\
  \hline
  B2 \\
  \cline{1-1}
  \hspace{-3em} r2  & H1-b & 35.6100 & -0.0346 & 0.56 (33) & 214$\pm$32  & 47.29 & B0V   & 0.068 & 0.16 & 2.04 & 1.51 \\
  \hspace{-3em} r3  & H1-c & 35.5895 & -0.0252 & 0.61 (36) & 474$\pm$36  & 47.63 & O9.5V & 0.088 & 0.15 & 2.20 & 1.40 \\
  \hspace{-3em} r4  & H2 & 35.5783 & -0.0303 & ... & 87$\pm$6    & 46.89 & B0.5V & 0.050 & 0.003 & 1.88 & 1.63 \\
  \hline
  B3 \\
  \cline{1-1}
  \hspace{-3em} r5  & H3 & 35.5735 & 0.0685  & 0.42 (24) & 420$\pm$15  & 47.58 & O9.5V & 0.085 & 0.08 & 2.17 & 1.42 \\
  \hspace{-3em} r6  & H4 & 35.5800 & 0.0657  & ... & 14$\pm$1    & 46.10 & B1.5V & 0.027 & 0.002 & 1.60 & 1.93 \\
  \hline
  B4 \\
  \cline{1-1}
  \hspace{-3em} r7  &  /  & 35.5777 & 0.0048  & ... & 53$\pm$2    & 46.68 & B1V   & 0.043 & 0.002 & 1.80 & 1.71 \\
  \hline
  B5 \\
  \cline{1-1}
  \hspace{-3em} r8  & H5   & 35.5437 & 0.0023  & 0.77 (45) & 209$\pm$30  & 47.28 & B0V   & 0.067 & 0.28 & 2.04 & 1.51 \\
  \hline
  B6 \\
  \cline{1-1}
  \hspace{-3em} r9  & H6   & 35.4728 & 0.0015  & 0.92 (54) & 104$\pm$89  & 46.97 & B0.5V & 0.053 & 0.45 & 1.91 & 1.61 \\
\enddata
\tablecomments{(1) Name of the radio subregions with their radio peaks. (2) Corresponding known \HII regions for radio peaks. (3)-(4) Galactic coordinates of the radio peaks. (5) Effective radius of the \HII regions. (6) Integrated flux density of the ionized gas at $\nu=1.4$ GHz (or MAGPIS 20 cm). (7) Hydrogen-ionizing photon rate. (8) Spectral type of the exciting O/B-type stars. (9) Radius of the Str$\ddot{\rm o}$mgren spheres. (10) Dynamical age of the \HII regions. (11)-(12) Radius and timescale when the \HII regions start to fragment. 
}
\end{deluxetable*}

\subsection{Young stellar population} \label{subsec:YSOs}

The preceding results establish N68 as an active massive star-forming region (MSFR). 
For comparison, we also investigate low-mass star formation by conducting a census of young stellar objects (YSOs). 
YSOs are typically classified as Class I, Flat-spectrum, Class II, or Class III based on their infrared spectral index or color excess, which traces the amount of circumstellar material. 
Following the methodologies outlined in \cite{Dewangan2018}, \cite{Baug2018}, and \cite{Chen2025}, we employed a multi-step identification scheme. 
Prior to classification, we retrieved near- to mid-infrared point-source catalogs covering a $18^{\prime} \times 18^{\prime}$ field centered on $(l, b) = (35\fdg58, -0\fdg03)$. 
The data include the UKIDSS $J$, $H$, and $K$ bands, the four $Spitzer$-IRAC bands ([3.6], [4.5], [5.8], and [8.0] $\mu$m), and the $Spitzer$-MIPS [24] $\mu$m band. 
Sources from these catalogs were cross-matched within a $1^{\prime\prime}$ radius to create a unified point-source catalog for subsequent YSO identification. 

\begin{figure}
    \centering
    \includegraphics[width=1\linewidth]{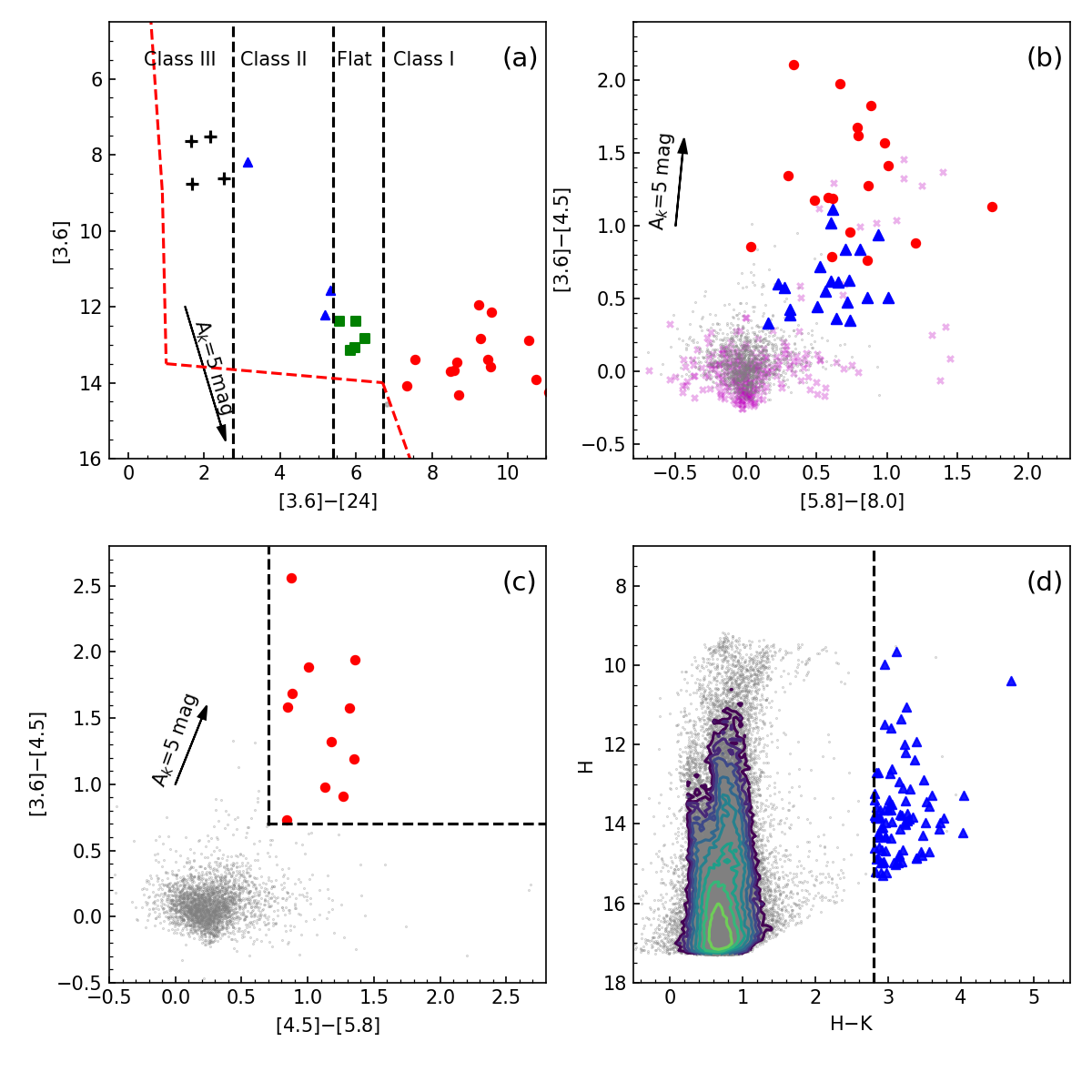}
    \caption{YSOs classification scheme based on the NIR/MIR color-magnitude diagrams (CMDs) and color-color diagrams (CCDs). Panel (a) presents the [3.6]–[24] vs. [3.6] color–magnitude diagram (CMD) employed in Method 1, with dashed lines separating Class I, Flat-spectrum, Class II, and Class III sources following \cite{Rebull2011}. Panel (b) shows the [5.8]–[8.0] vs. [3.6]–[4.5] color–color diagram (CCD) for sources detected in all four IRAC bands, with contaminants removed using the criteria of \cite{Gutermuth2009}. Panel (c) displays the [4.5]–[5.8] vs. [3.6]–[4.5] CCD for sources detected in the first three IRAC bands, applying the selection thresholds from \cite{Hartmann2005} and \cite{Getman2007}. Panel (d) presents the H–K vs. K CMD for sources only detected in the H and K bands; a cutoff of H–K > 2.8 mag, derived from a nearby control field, is used to identify heavily reddened Class II candidates.}
    \label{fig:YSOs}
\end{figure}

(i) CMD $[3.6]-[24]/[3.6]$: Band [24] reveals the emission from the protostars deeply embedded in the molecular cloud cores. 
Therefore, using the reddening of color [3.6]$-$[24], protostars can be effectively separated. 
Following the classification criteria established by \cite{Guieu2010} and \cite{Rebull2011}, we separate YSOs into Class I, Flat-spectrum, Class II, and Class III sources based on different reddening. 
This method yields 16 Class I, 5 Flat-spectrum, 3 Class II, and 4 Class III candidates, as shown in Figure \ref{fig:YSOs}(a). 

(ii) CCDs among bands [3.6], [4.5], [5.8] and [8.0]: For sources detected in all four IRAC bands, we constructed a series of color–color diagrams following the methodology outlined in Phase 1 of \cite{Gutermuth2009}. This procedure effectively removes contaminants such as star-forming galaxies (SFGs), active galactic nuclei (AGNs), PAH-dominated sources, and shock knots. After excluding non-YSO sources, the remaining objects were classified as Class I or Class II based on established color criteria. Using this method, we identify 18 Class I and 21 Class II sources. The corresponding CCDs and classification results are presented in Figure \ref{fig:YSOs}(b). 

(iii) CCD $[3.6]-[4.5]/[4.5]-[5.8]$: This method is an improvement on method (ii), which omits the stars in band [8.0] to reduce the contamination of PAH emissions. Using the simple criteria \citep{Hartmann2005, Getman2007, Baug2018} of $[4.5]-[5.8]\geq0.7$and $[3.6]-[4.5]\geq0.7$, we obtain a total of 11 Class I YSOs and show the results in Figure \ref{fig:YSOs}(c). 

(iv) CMD ${\rm H-K/K}$: This method utilizes the high sensitivity of the $H$ and $K$ near-infrared bands to detect faint, heavily reddened YSOs. 
A color cutoff of ${\rm H-K} = 2.8$ was adopted, determined from the color–magnitude distribution of field stars in a nearby control region. 
Sources satisfying ${\rm H-K} > 2.8$ are classified as candidate Class II objects. 
In total, 89 such reddened sources were identified (see Figure \ref{fig:YSOs}(d)). 

Based on a comprehensive multi-wavelength identification scheme, we have cataloged a total of 163 YSOs in the R1 region of N68. 
This population includes 45 Class I sources, 5 Flat-spectrum sources, and 113 Class II sources, with an additional 4 Class III sources identified. 
The significant number of Class I and Class II objects confirms that both embedded and more evolved low- to intermediate-mass star formation is actively ongoing in this region.

\subsection{Star formation and distribution} \label{subsec:SF}

\begin{figure}
    \centering
    \includegraphics[width=1\linewidth]{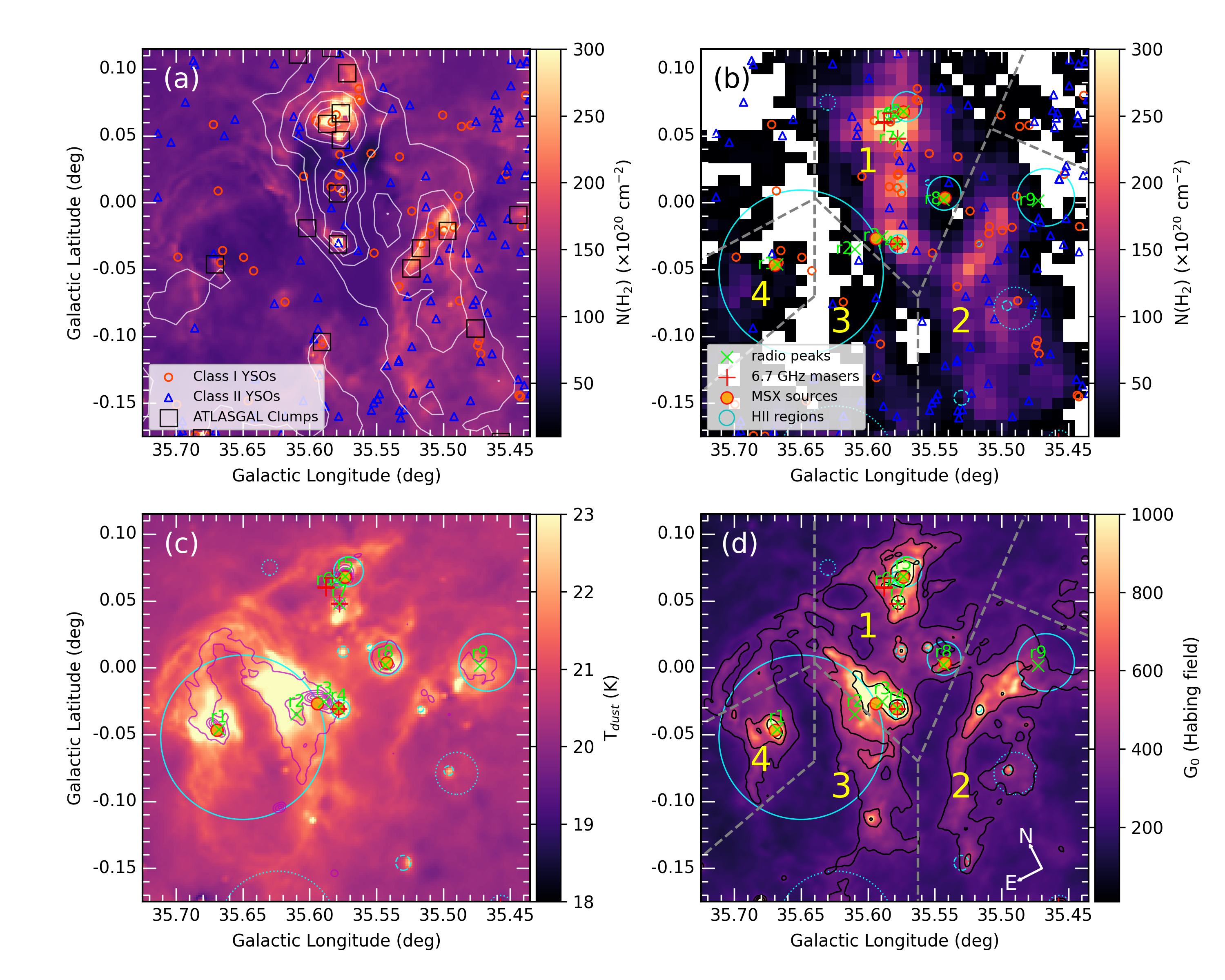}
    \caption{Panel (a) Column density extracted from PPMAP. The white dotted contours indicate the $^{13}$CO moment 0 map integrated from 47 to 64 \kms with contour levels from 16 (3$\sigma$) to 56 K \kms in steps of 8 K \kms. The red open circles and black open triangles indicate the Class I and Class II YSOs identified in Section \ref{subsec:YSOs}, respectively. The black squares indicate the ATLASGAL clumps. Panel (b) Column density estimated from $^{13}$CO. The massive star formation signatures (6.7 GHz masers, $MSX$ sources and \HII regions) are drawn as the same manners as in Figure \ref{fig:overlay}. Panel (c) Dust temperature extracted from PPMAP. Panel (d) FUV field. The contour levels at $G_0=$ 300, 500, 1000 and 3000 are drawn. The dashed lines in panels (b) \& (d) divided the R1 region into four main clouds, namely Clouds 1-4. }
    \label{fig:CTG}
\end{figure}

Figures \ref{fig:CTG} (a) and (b) show the spatial distribution of the identified YSOs overlayed on the column density map derived from the PPMAP\footnote{\url{http://www.astro.cardiff.ac.uk/research/ViaLactea/PPMAP_Results/}} results of \cite{Marsh2017} and $^{13}$CO, respectively. 
As seen in Figure \ref{fig:CTG}(a), Class I YSOs are predominantly located along the dense ridges of the molecular cloud, particularly coincident with ATLASGAL clumps, indicating that these young protostars remain deeply embedded within their natal gas. 
In contrast, Class II YSOs are more widely dispersed around these dense ridges, suggesting that they have begun to drift away from or partially dispersed their immediate birth environments. 
Furthermore, the dust temperature map (Figure \ref{fig:CTG}(c)) reveals that regions with elevated temperatures ($>22$ K) closely correspond to the positions of known \HII regions, implying that the heating in these areas is dominated by radiation from embedded massive stars rather than by the collective emission of low-mass YSOs. 
We also derived the column density and mass of molecular gas using our CO data for comparison. The calculation is based on the local thermal equilibrium (LTE) assumption, and for more details, please refer to our previous work (\citealt{Chen2025}). 
Although the low-resolution CO data show fewer details, it is still clearly observable that the N68 molecular cloud is clumpy. 
By combining its spatial distribution, kinematic characteristics, and radiation field features (see below), we divided the N68 molecular cloud into four subregions, namely Clouds 1-4. 
Assuming that the YSOs projected on each cloud are formed by that cloud, we can estimate the stellar mass according to \cite{Chen2024a} using the assumptions of the disk fraction of 0.5, and the average mass of YSOs and MYSOs/OB-type stars of 0.64 M$_{\odot}$ and 10 M$_{\odot}$, respectively. 
And finally calculate the star formation efficiency based on the formula ${\rm SFE = M_{\star}/(M_{gas}+M_{\star})}$. 
The calculation results are listed in Table \ref{tab:clouds}. 
The SFE of these four clouds ranges from 0.8\% to 2.1\%, which is relatively smaller compared to the nearby clouds (i.e., 3.0-6.3\% by \citealt{Evans2009}).

\begin{deluxetable*}{ccccccccccccc}
    \tablecaption{Properties of clouds that induced star formation as classified in Figure \ref{fig:CTG}.  \label{tab:clouds}}
    \tablehead{
    \colhead{Cloud} & \colhead{$\overline{l}$} & \colhead{$\overline{b}$} & \colhead{R$_{eff}$} & \colhead{$\overline{N(H_2)}$} & \colhead{$N_{peak}$}  & \colhead{M$_{gas}$} & \colhead{$\overline{G_0}$} & \colhead{N$_{\rm YSO}$} & \colhead{N$_{\rm MYSO}$} & \colhead{N$_{\rm OB}$} & \colhead{M$_\star$}  & \colhead{SFE} \\
    \cline{1-13}
    \colhead{ID} & \colhead{($^{\circ}$)} & \colhead{($^{\circ}$)} & \colhead{($^{\prime}$)} & \colhead{($10^{20}$ cm$^{-2}$)} & \colhead{($10^{20}$ cm$^{-2}$)} & \colhead{($10^3$ M$_{\odot}$)} & \colhead{(Habing field)} & \colhead{(I+II)} & \colhead{} & \colhead{} & \colhead{(M$_{\odot}$)} & \colhead{($\%$)}
    }
    \colnumbers
    \startdata
        1 & 35.583 & 0.045 & 4.39 & 91 & 358 & 12 & 500 & 16+16 & 4 & 6 & 131 & 1.0\\
        2 & 35.493 & -0.072 & 4.84 & 59 & 220 & 9.4 & 281 & 16+44 & 0 & 1 & 77 & 0.8\\
        3 & 34.641 & -0.108 & 3.91 & 23 & 59 & 2.4 & 257 & 6+16 & 0 & 1 & 34 & 1.4\\
        4 & 35.686 & -0.070 & 2.69 & 24 & 76 & 1.2 & 280 & 5+2 & 1 & 1 & 26 & 2.1 \\
    \enddata
    \tablecomments{(1) IDs for the clouds. (2)-(3) Galactic coordinates of geometric center of the clouds. (4) Effective radius. (5)-(6) Average and peak column density. (7) Gas mass. (8) Average intensity of FUV field. (9)-(11) Number of YSO, MYSO and OB star detected in each cloud. (12) Stellar mass estimated in each cloud. (13) Star formation efficiency of each cloud.  }
\end{deluxetable*}

Furthermore, in order to detect the irradiated structures of the molecular cloud, we also derive the local far-ultraviolet (FUV) radiation field from the observed intensity of far-infrared (FIR) radiation in our $Herschel$ maps. 
The approach is described in more detail by \cite{Kramer2008}, 
\begin{equation}
    G_0=\frac{4\pi\ I_{FIR}}{1.6\times10^{-3}\ {\rm erg\ cm^{-2}\ s^{-1}}}, 
\end{equation}
The unit of the FUV field is the so-called Habing field, i.e., the average intensity of the local Galactic FUV ($912-2400\ \text{\AA}$) flux of $1.6\times10^{-3}\ {\rm erg\ cm^{-2}\ s^{-1}}$ \citep{Habing1968}. 
According to \cite{Roccatagliata2013}, the total $60-200$ $\mu$m FIR intensity, $I_{FIR}$, can be derived from the 70 and 160 $\mu$m maps, 
\begin{equation}
    I_{FIR}=f_{70}\cdot \Delta\lambda_{70} + \frac{1}{2}(f_{70}+f_{160})\cdot \Delta\lambda_{100} + f_{160}\cdot \Delta\lambda_{160}, 
\end{equation}
Where $f_{70}$ and $f_{160}$ are the flux densities observed in 70 and 160 $\mu$m bands, respectively. 
And $\Delta\lambda_{70}=20\ \mu$m ($60-80\ \mu$m), $\Delta\lambda_{100}=45\ \mu$m ($80-125\ \mu$m), $\Delta\lambda_{160}=75\ \mu$m ($125-200\ \mu$m) are the bandwidths for the filters at 70, 100, and 160 $\mu$m in $Herschel$ PACS MANUAL\footnote{\url{http://herschel.esac.esa.int/Docs/PACS/html/ch03s02.html}}. 
Due to the lack of observation at the 100 $\mu$m band, here we simply used the average values of the 70 and 160 $\mu$m intensity to replace it. 
Note that before running the calculation, the PACS 70 $\mu$m map (5.8$^{\prime\prime}$) should first smooth to the angular resolution of the PACS 160 $\mu$m map (12$^{\prime\prime}$) and regrid to the same pixel size (i.e., 3.2$^{\prime\prime}$).

Through the calculations mentioned above, we obtain the FUV field map of N68 as shown in Figure \ref{fig:CTG}(d). 
The contour levels at $G_0=$ 300, 500, 1000 and 3000 are drawn in the figure. 
Typically, for regions containing multiple O/B-type stars, molecular cloud surfaces are irradiated with $G_0$ values above $\sim300$ \citep{Hollenbach1991}. 
Similar to Figure \ref{fig:CTG} (b), the FUV field is mainly distributed in the four subregions of Clouds 1-4. 
The average $G_0$ value of Cloud 1 is 500, while the average $G_0$ values of Clouds 2-4 are less than 300, indicating that the radiation field of Clouds 2-4 is affected by the radiation sources in Cloud 1 or other external ionized sources. 
This is consistent with the fact that Clouds 3 and 4 are located at the edge of the N68 \HII region, and the radiation of Cloud 2 has an arc-shaped structure pointing to Cloud 1.

By comparing the dust temperature map with the FUV intensity map, their distributions are closely related. 
In the vicinity of massive stars or in the shell of \HII regions, both dust temperature and FUV intensity are relatively high, indicating their irradiation origin. 
However, their differences still exist. 
For instance, the more diffuse, low-density dust in the Central region is readily heated by radiation, leading to a spatially extended warm dust distribution, while the FUV field peaks more sharply at the locations of the massive stars themselves, resulting in a slight offset between the two maps. 
Overall, the observed enhancements in both dust temperature and FUV field across the N68 molecular cloud are primarily driven by the combined radiation from multiple embedded O/B-type stars.

\subsection{Molecular structure and kinematics} \label{subsec:structure}

\begin{figure}
	\includegraphics[width=0.337\textwidth]{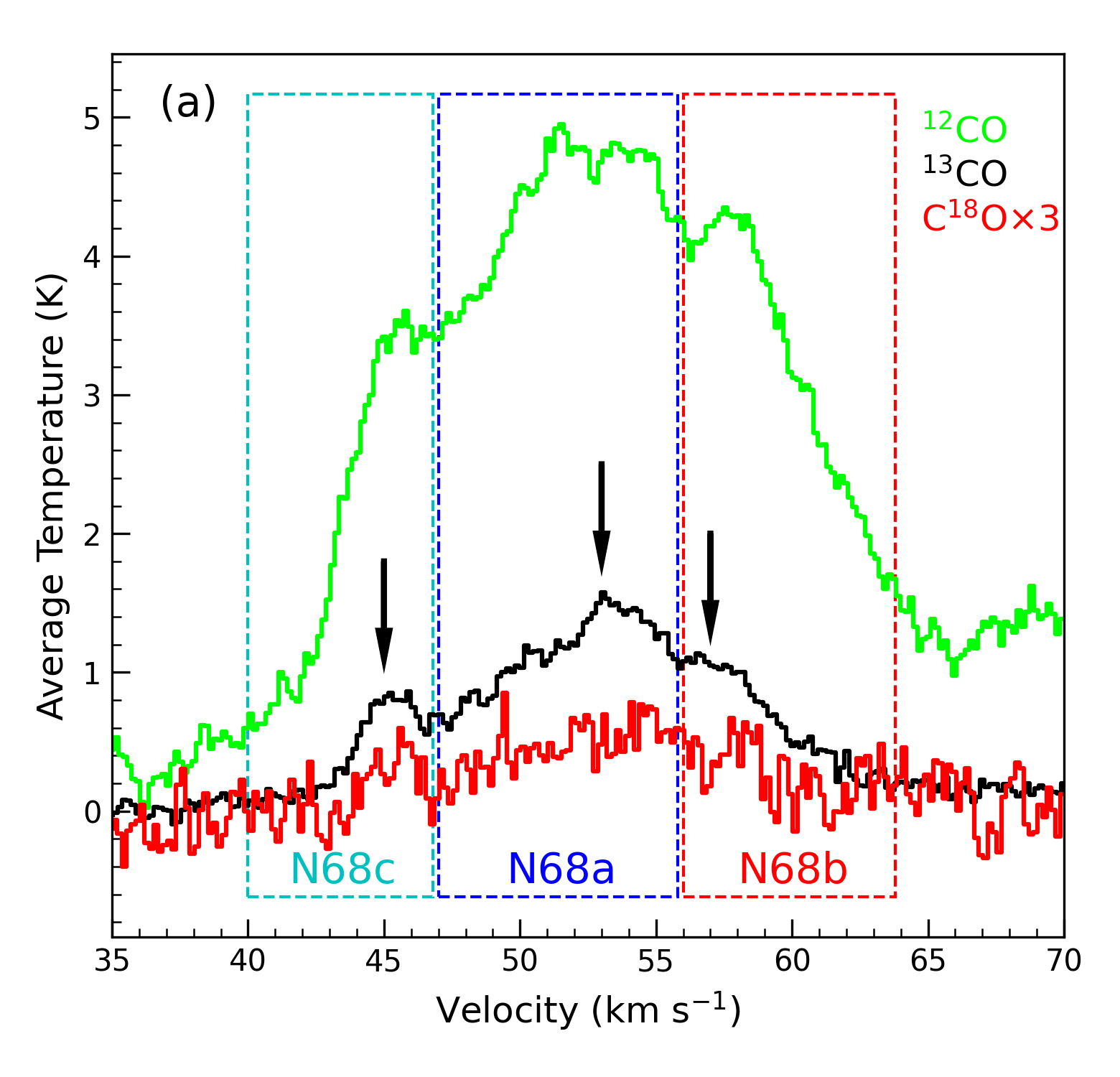}
    \includegraphics[width=0.663\textwidth]{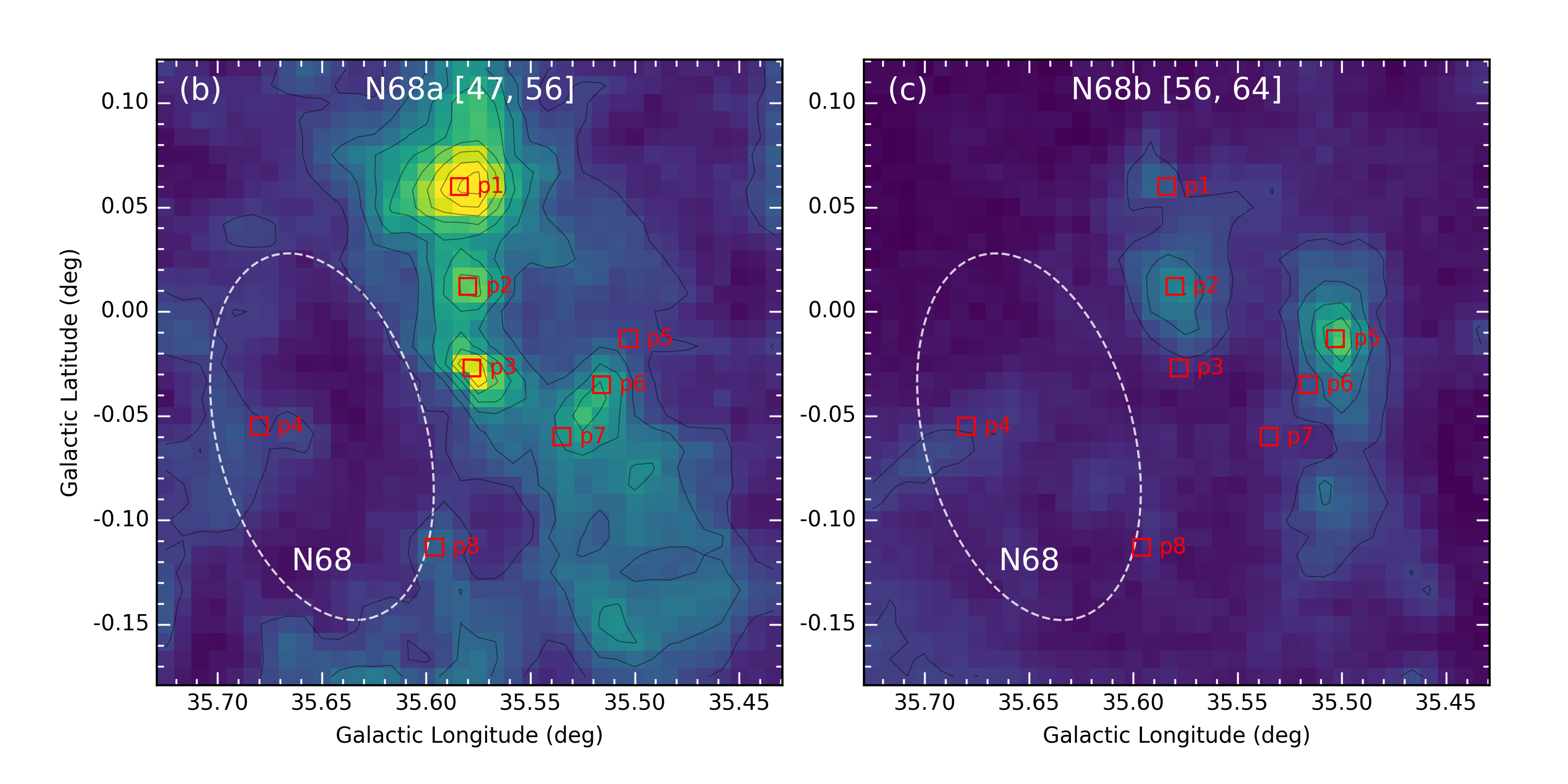}
    \caption{Overview of the average spectra and spatial distribution of the main components related to bubble N68. Panel (a) shows the average spectra of the R1 region. Green, black and red lines indicate the average spectra of $^{12}$CO, $^{13}$CO, and C$^{18}$O, respectively. Blue, red and cyan dashed boxes highlight the three main gas components, namely N68a [47, 56], N68b [56, 64] and N68c [40, 47], respectively. Black arrows depict the system velocity for each components (i.e., 53, 57 and 45 \kms for N68a, N68b and N68c, respectively). Panels (b) and (c) show the $^{13}$CO spatial distribution of N68a (integrated from 47 to 56 \kms) and N68b (integrated from 56 to 64 \kms), respectively. The contours are drawn at 8.1 K \kms (3$\sigma$) to 45.9 K \kms in steps of 5.4 K \kms (2$\sigma$). The red squares labeled p1 to p8 represent the positions of the local intensity peaks, where the spectra are extracted in Figure \ref{fig:spectra}. }
    \label{fig:m0}
\end{figure}

We characterize the molecular structure and kinematics of the N68 region using the three CO isotope spectral lines ($^{12}$CO, $^{13}$CO and C$^{18}$O $J=1-0$) observed with the PMO-13.7m telescope. 
Figure \ref{fig:m0}(a) displays the spatially averaged spectra over the R1 region for $^{12}$CO (green), $^{13}$CO (black), and C$^{18}$O (red). 
Three distinct velocity components are evident: N68a, N68b, and N68c, with systemic velocities around 53, 57, and 45 \kms, respectively. 
The components N68a (47–56 \kms) and N68b (56–64 \kms) are spectrally adjacent and both have clear infrared counterparts, confirming their association with bubble N68. 
In contrast, component N68c (40–47 \kms), although spectrally contiguous with N68a, lacks an obvious infrared counterpart and is therefore not discussed further in this work. 
Figures \ref{fig:m0}(b) and (c) present the $^{13}$CO integrated intensity (moment 0) maps for the N68a and N68b components, respectively. 
Both clouds exhibit clumpy structures, with local intensity peaks labeled p1$-$p8. 
Peaks p1, p3, p4, p7, and p8 are associated with the N68a cloud, while p5 belongs to N68b. 
Notably, peaks p2 and p6 lie at the interface where the two clouds overlap, appearing in both components. 
The morphology of N68a is largely shaped by the expansion of bubble N68. 
Gas has been swept up and accumulated along the expanding shell of the \HII region, forming the characteristic bubble structure. 
This is supported by the location of the peaks p3, p4, and p8 along the molecular shell of the bubble. 
In contrast, the N68b cloud is located farther from the bubble and shows less morphological disturbance from its expansion.

\begin{figure}
	\includegraphics[width=1\textwidth]{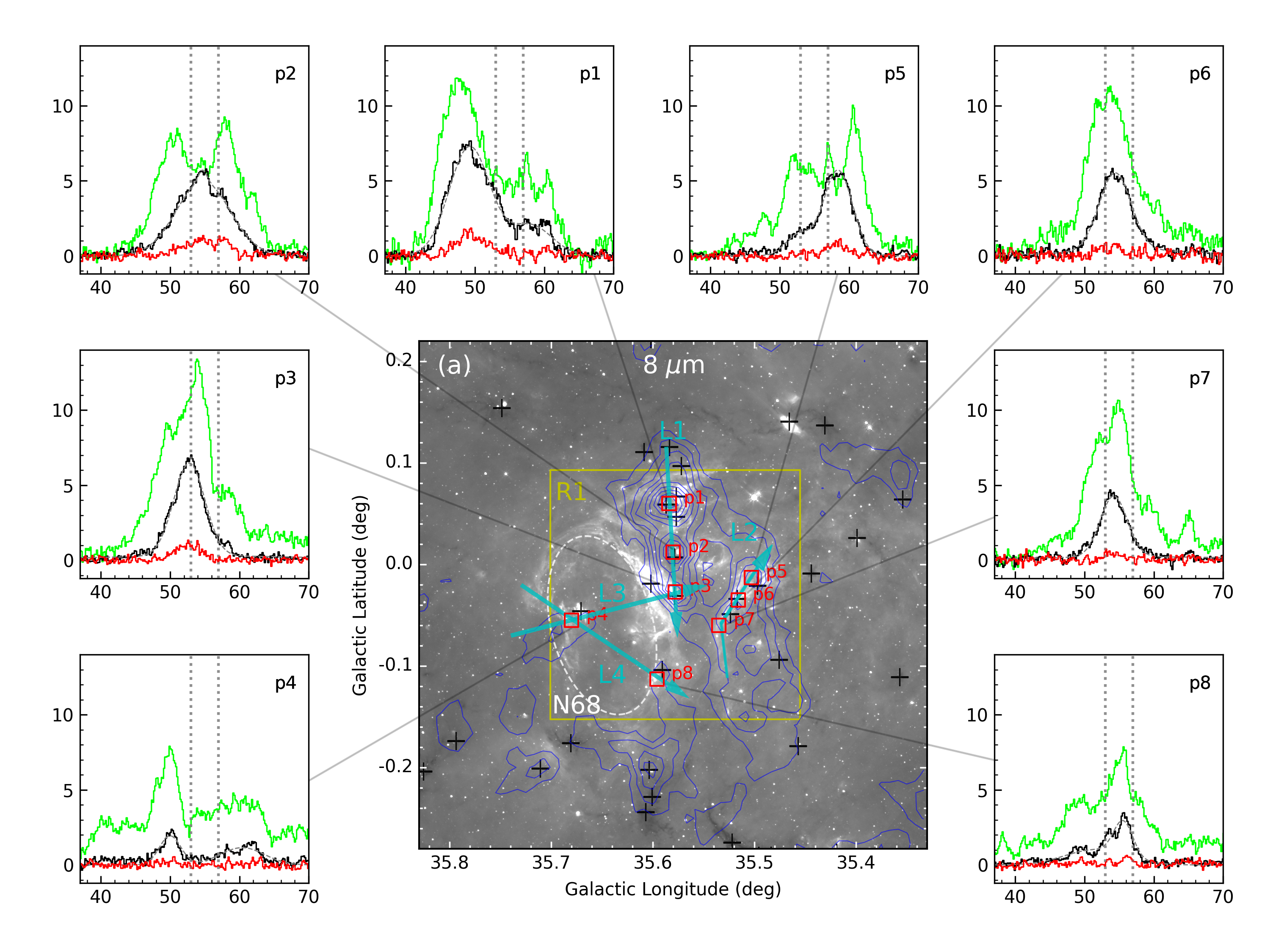}
    \includegraphics[width=0.245\textwidth]{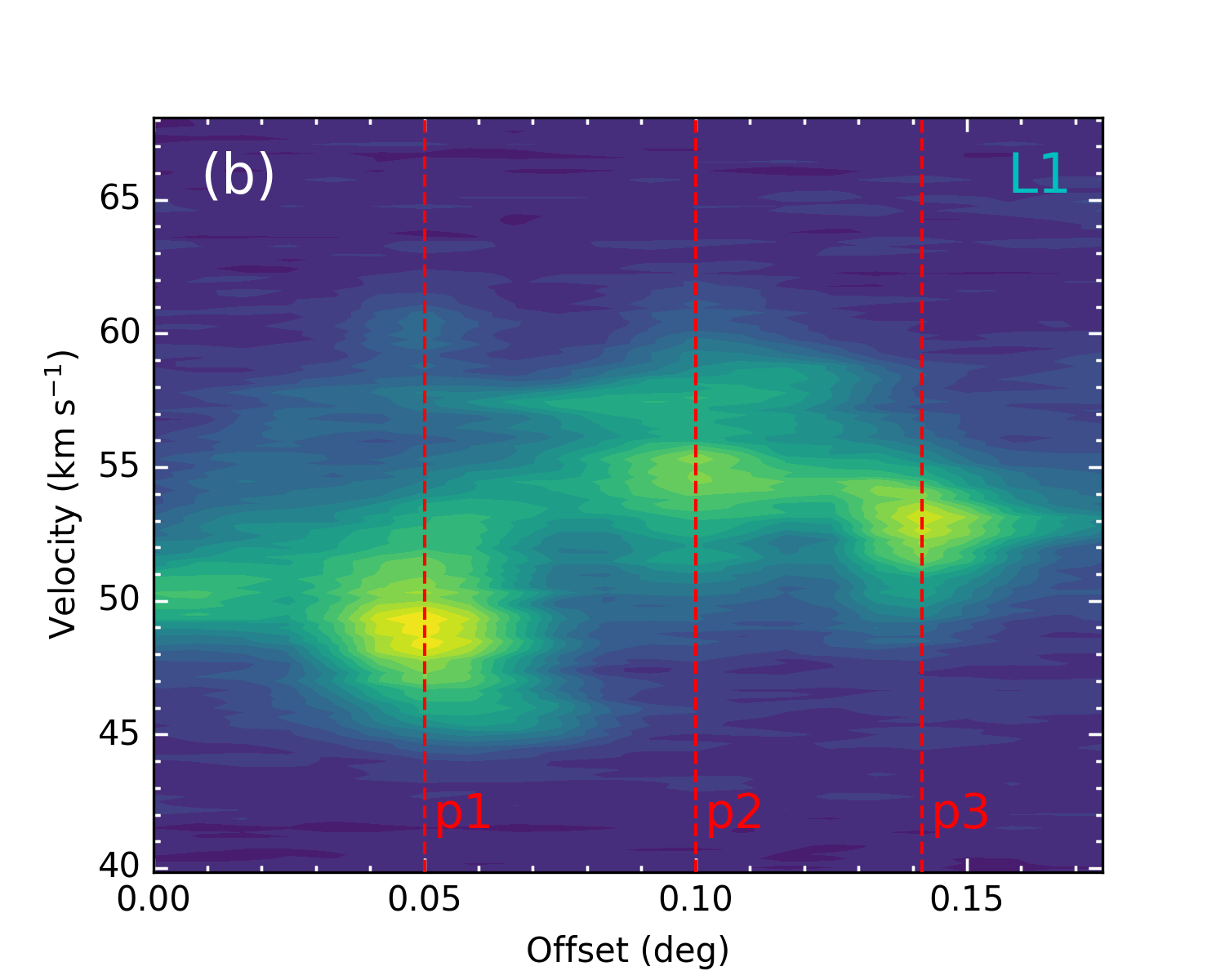}
    \includegraphics[width=0.245\textwidth]{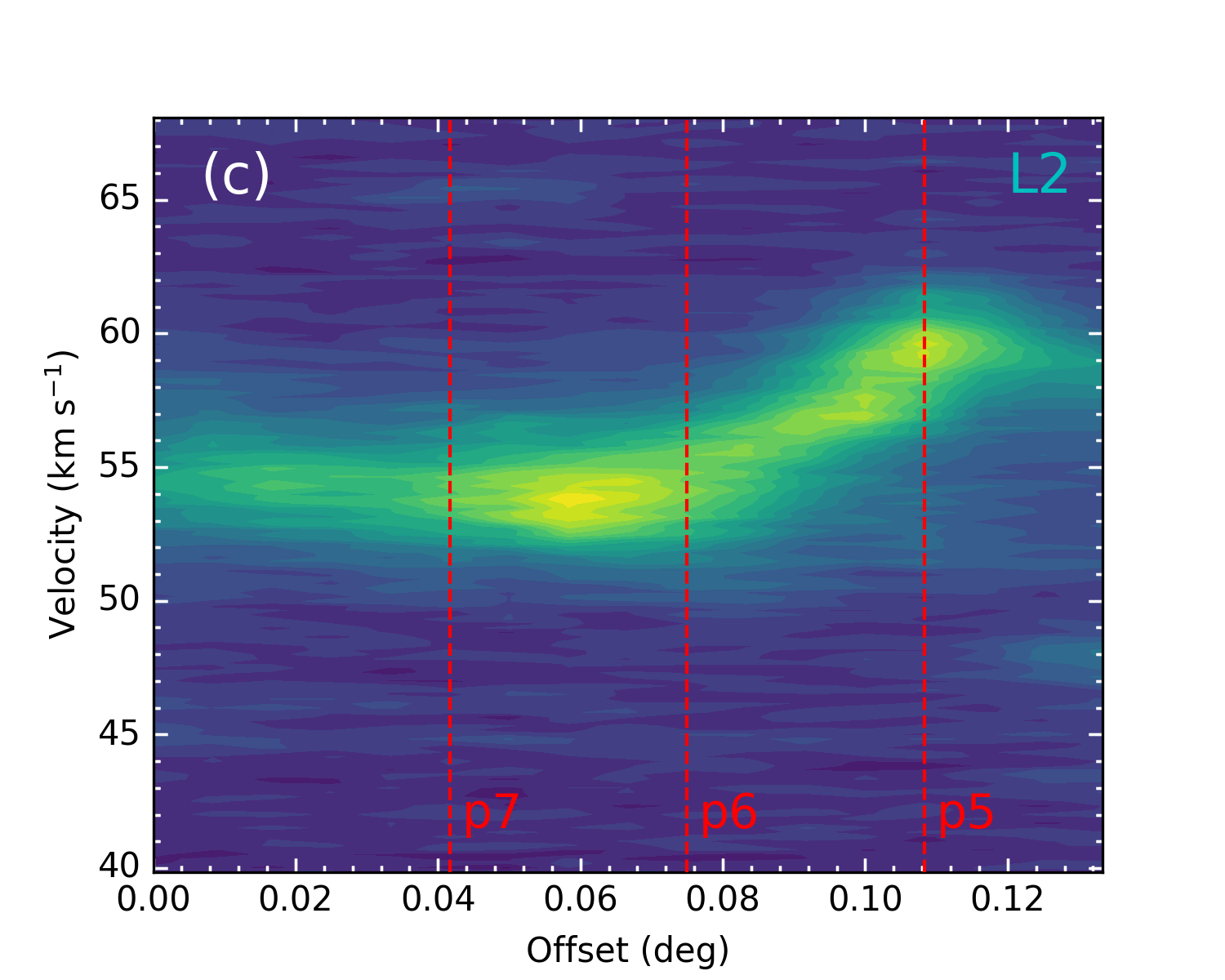}
    \includegraphics[width=0.245\textwidth]{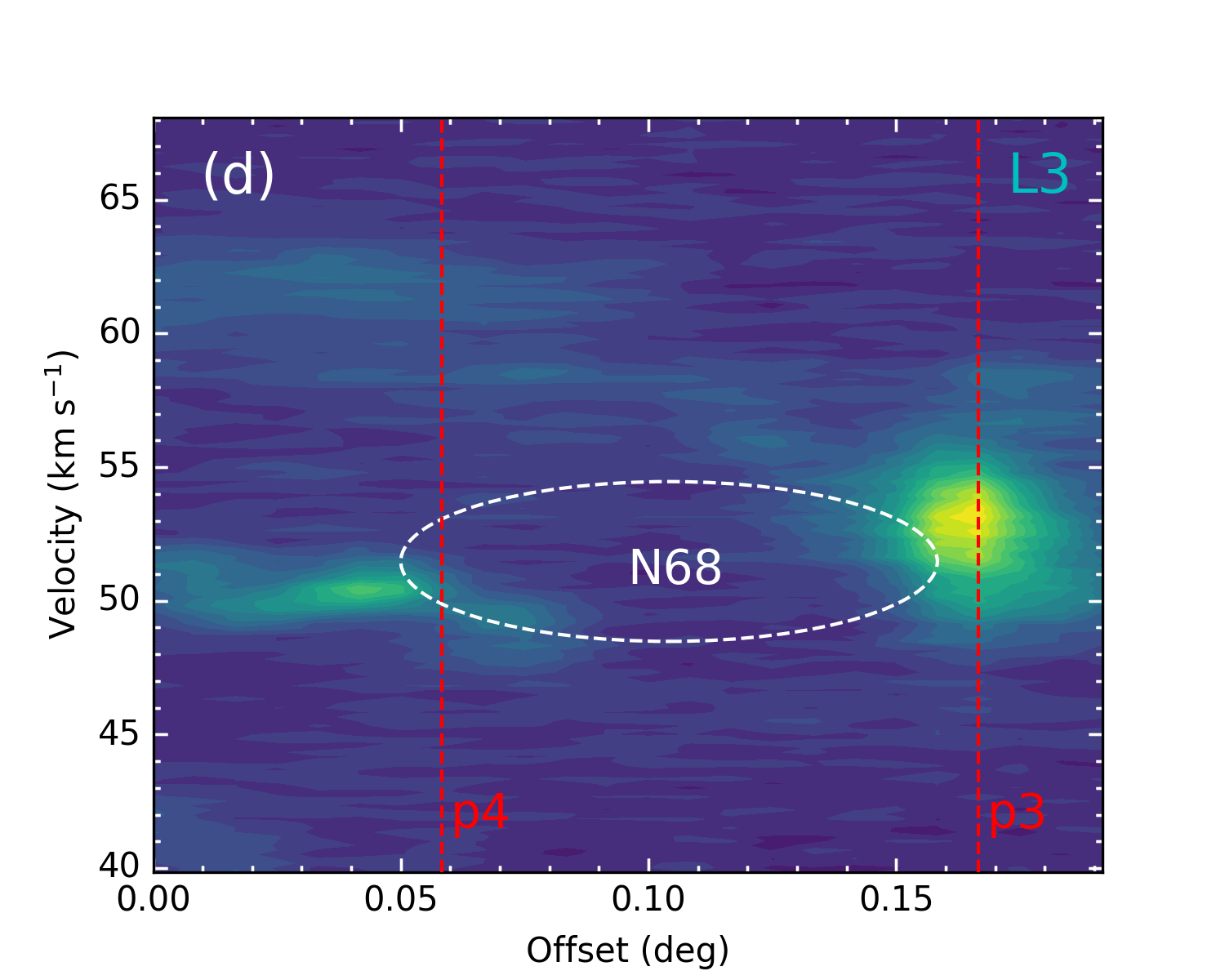}
    \includegraphics[width=0.245\textwidth]{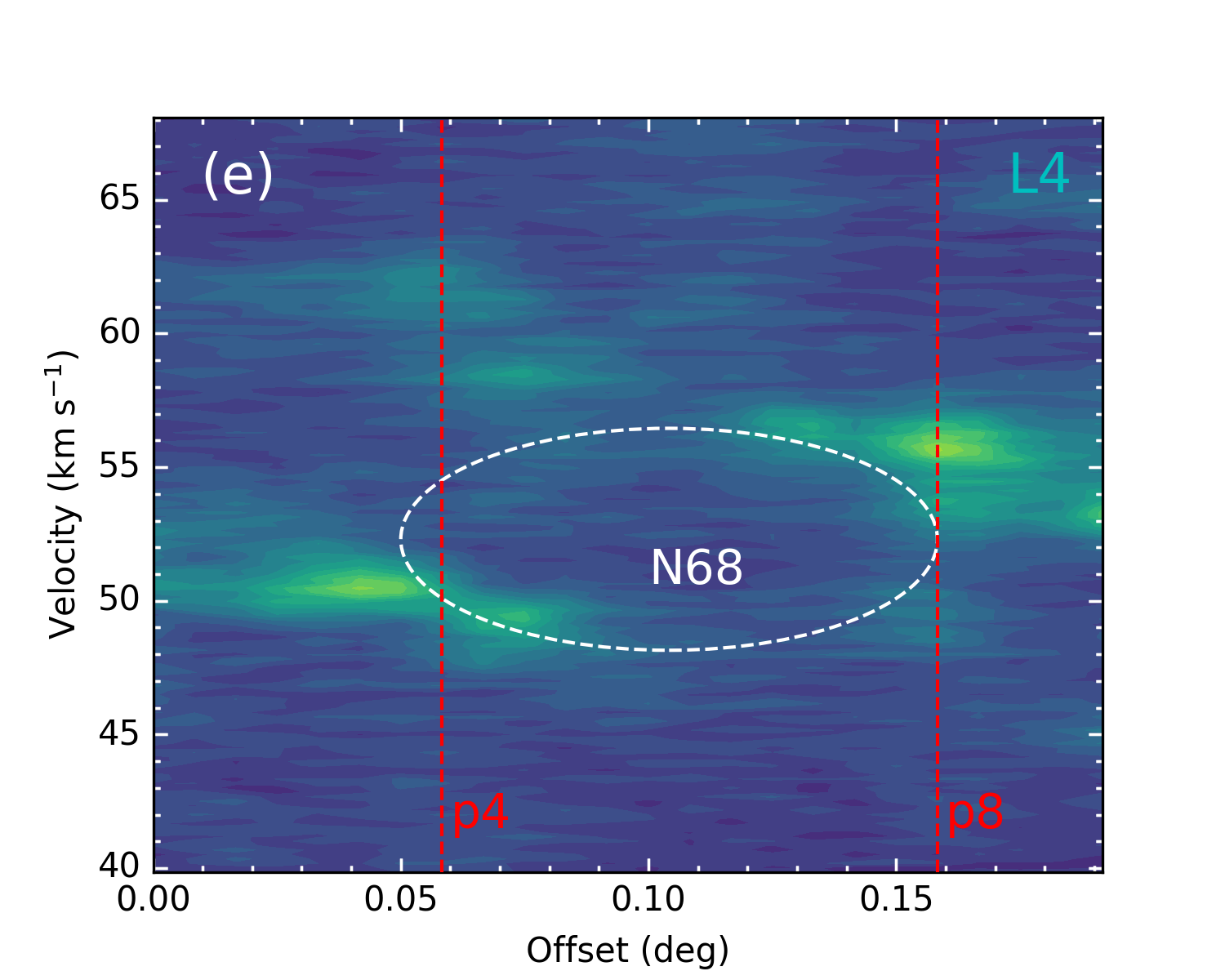}
    \caption{Spectra and PV slices of N68. Panel (a) shows the integrated intensity $^{13}$CO (1-0) map of the region and the locations for extracting spectra. The contour levels start from 15 (3$\sigma$) to 55 K \kms in steps of 5 K \kms (1$\sigma$). The background image shows the 8 $\mu$m emission. The yellow box highlights the region of R1. The white dashed ellipse depicts the layout of the N68 bubble. The black pluses represent the ATLASGAL clumps. Local integrated intensity peaks have been marked by red squares and labeled p1 to p8. $^{12}$CO (1-0) (green), $^{13}$CO (1-0) (black) and C$^{18}$O (1-0) (red) spectra at each of these peak locations has been extracted. The black dashed lines are Gaussian fitting lines for $^{13}$CO spectra. Panels (b) to (e) show the PV slices along the paths L1 to L4 as shown in panel (a). The vertical dashed lines indicate the locations of local peaks on each path. }
    \label{fig:spectra}
\end{figure}

To examine the kinematic structure of the bubble and molecular clouds in more detail, we extracted the spectral lines ($^{12}$CO, $^{13}$CO and C$^{18}$O), as well as position–velocity (PV) diagrams, along the peaks p1$-$p8 (Figure \ref{fig:spectra}). 
For each position, we performed Gaussian fitting on the $^{13}$CO spectral lines to characterize the velocity components. 
The fitting parameters for both main and secondary components are listed in Table \ref{tab:gaussian}. 
As mentioned above, p1, p3, p4, p7, and p8 are associated with the N68a cloud, though their spectral profiles show distinct kinematic signatures. 
The spectrum at p3 is well fitted by a single Gaussian centered at 52.8 \kms. 
In contrast, spectra at p1 and p4 display a significant blueshifted, with peak velocities around 50 \kms, while p7 and p8 are slightly redshifted, with peaks between 54 and 56 \kms. 
These systematic velocity shifts can be attributed to the expansion of bubble N68, which imparts different line-of-sight motions to gas along the front (blueshifted) and back (redshifted) sides of the shell. 
As shown in the PV slices across the bubble (Figures \ref{fig:spectra}(d) and (e); cuts p4–p3 and p4–p8), a clear pattern of redshifted and blueshifted emission is observed. 
The velocity separation between these features reflects the expansion motion of the shell, from which we estimate an expansion velocity of $3\sim5$ \kms for bubble N68. 
Interestingly, the spectrum at position p4 shows a slight velocity gradient. 
Given that p4 coincides with the bright rim cloud (BRC), this gradient may arise from the differential motion between the original velocity of the pre-existing dense core and the velocity imposed by the expanding \HII region.

The peak p5, located within the N68b molecular cloud, exhibits a peak velocity of 58.4 \kms, which is significant different from the velocity change caused by the bubble expansion (50-56 \kms). 
This clear kinematic separation confirms that N68a and N68b are two relatively independent molecular structures. 
In contrast, the spectra at positions p2 and p6 display single, broad Gaussian profiles centered at $\sim55$ \kms, which is an intermediate velocity between the systemic velocities of N68a and N68b. 
This suggests that the two clouds spatially connect or overlap at these locations. 
The PV slices across p2 and p6 (Figures \ref{fig:spectra}(b) and (c)) further reveal that these points mark the transition or junction between the N68a and N68b velocity components. 
Such kinematic signatures, with broad intermediate-velocity features at cloud interfaces, are characteristic of cloud–cloud collisions. 
We therefore infer that a CCC process is likely occurring at p2 and p6. 
A more detailed analysis of this interaction is presented in Section \ref{subsec:bridge}.

\begin{deluxetable*}{ccccccccc}
    \tablecaption{Gaussian fitting parameters of the local $^{13}$CO peaks.  \label{tab:gaussian}}
    \tablehead{
    \colhead{Peak} & \colhead{$l$} & \colhead{$b$} & \colhead{T$_1$} & \colhead{$v_1$} & \colhead{$\Delta v_1$}  & \colhead{T$_2$} & \colhead{$v_2$} & \colhead{$\Delta v_2$} \\
    \cline{1-9}
    \colhead{ID} & \colhead{($^{\circ}$)} & \colhead{($^{\circ}$)} & \colhead{(K)} & \colhead{(\kms)} & \colhead{(\kms)}  & \colhead{(K)} & \colhead{(\kms)} & \colhead{(\kms)} 
    }
    \colnumbers
    \startdata
        p1 & 35.584 & 0.060 & 7.3 & 49.2 & 6.9 & 2.0 & 58.0 & 7.3\\
        p2 & 35.580 & 0.012 & 5.0 & 54.8 & 9.2 & ... & ... & ...\\
        p3 & 35.578 & -0.027 & 6.2 & 52.8 & 5.3 & ... & ... & ...\\
        p4 & 35.602 & -0.202 & 1.9 & 50.0 & 3.5 & 1.2 & 60.4 & 7.5 \\
        p5 & 35.503 & -0.013 & 5.7 & 58.4 & 4.9 & 1.4 & 52.7 & 4.4 \\
        p6 & 35.517 & -0.043 & 5.2 & 54.4 & 5.4 & ... & ... & ... \\
        p7 & 35.505 & -0.083 & 4.1 & 54.2 & 4.5 & ... & ... & ... \\
        p8 & 35.511 & -0.202 & 2.3 & 55.6 & 3.3 & 1.0 & 49.5 & 2.0 \\
    \enddata
    \tablecomments{(1) IDs for the peaks. (2)-(3) Galactic coordinates of the peaks. (4)-(6) Fitting parameters (intensity, mean velocity and line width of the $^{13}$CO spectral lines, respectively) of the main component. (7)-(9) Fitting parameters of the secondary component. }
\end{deluxetable*}

\section{Discussion} \label{sec:discussion}

\subsection{Signature of CCC} \label{subsec:CCC}

The review by \cite{Fukui2021} summarizes several key observational signatures of CCC, including: i) the broad bridge feature; ii) the spatial complementary distribution, and iii) the U-shaped cavity structure. 
Identification of CCCs therefore relies on matching these morphological and kinematic signatures in molecular gas observations. 
This approach has been successfully applied to confirm CCCs in numerous massive star forming regions, such as M 20 \citep{Torii2011}, RCW 120 \citep{Torii2015}, M 42 \citep{Fukui2018b}, and W 43 \citep{Kohno2021}, etc. 
In the following analysis, we demonstrate that the N68 region exhibits clear signatures consistent with an ongoing CCC process.

\subsubsection{Broad bridge feature} \label{subsec:bridge}

The presence of a broad bridge feature in position-velocity phase is a key kinematic signature indicating physical contact between two molecular clouds. 
This intermediate-velocity gas arises from the deceleration and mixing of material during the collision. 
As the kinetic motion of the clouds is converted into turbulent motion, the velocity dispersion increases near the interface \citep{Fukui2018b}. 
The broad bridge is a transient feature; its lifetime is influenced by stellar feedback, star formation efficiency, and the duration of the collision itself. 
According to \cite{Haworth2015a}, such bridges can remain detectable for up to $\sim$2.5 Myr after the onset of massive star formation before radiative feedback significantly disrupts them. 
Subsequent simulations by \cite{Haworth2015b} suggest that the bridge lifetime is primarily set by the collision timescale rather than by feedback disruption alone. 
In N68, a broad bridge feature is clearly detected at positions p2 and p6. 
As shown by the single Gaussian fits in Figure \ref{fig:spectra}, the $^{13}$CO line widths at these locations are 9.2 and 5.4 \kms, respectively. 
These widths are approximately twice the typical line width ($\sim3$ \kms) observed in the undisturbed portions of N68a or N68b. 
Although the N68 \HII region (H1) has a dynamical age of 2.9 Myr, its direct influence on the kinematics at p2 and p6 appears limited, as these collision sites are not spatially associated with the ionized region. 
Therefore, the significantly broadened lines are more likely to result from turbulent injection from a cloud–cloud collision rather than expansion or feedback from the nearby \HII region. 

\begin{figure}
    \centering
    \includegraphics[width=1\textwidth]{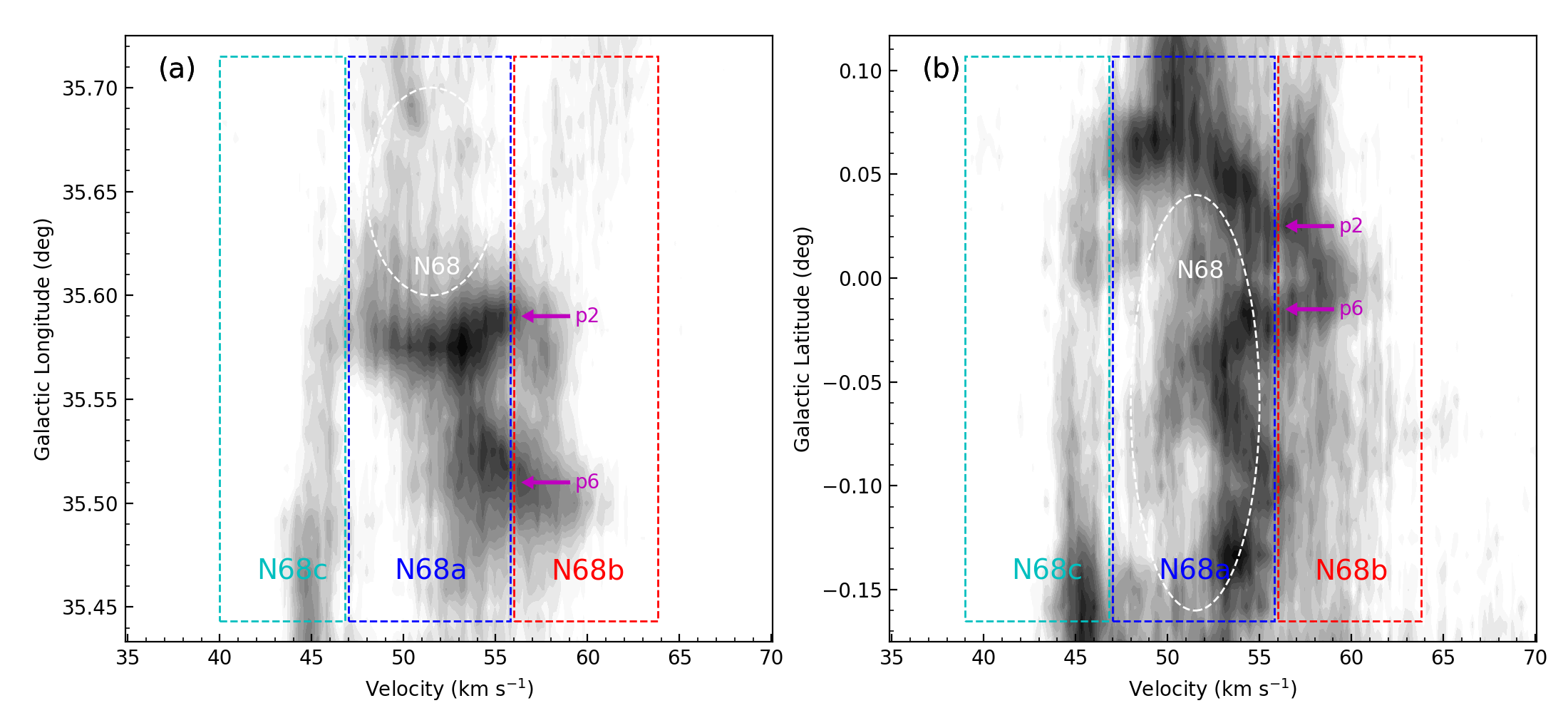}
    \caption{Position-Velocity (PV) maps of $^{13}$CO (1-0) gas of the R1 region. Panel (a) and (b) are $l$-PV and $b$-PV maps, respectively. Three prominent gas components, namely N68a, N68b and N68c, are highlighted by blue, red and cyan dashed boxes, respectively. 
    The projection of bubble N68 is depicted by white dashed ellipse. The bridge features connecting N68a and N68b are marked by magenta arrows, which correspond to the locations of p2 and p6 in Figure \ref{fig:spectra}, respectively. }
    \label{fig:pv}
\end{figure}

Figure \ref{fig:pv} presents the $^{13}$CO PV diagram of the R1 region. 
Two relatively independent molecular clouds are evident: N68a, with a systemic velocity around 53 \kms, and N68b, centered near 57 \kms. 
These components are connected by broad bridge features at positions p2 and p6 (depicted by magenta arrows). 
The PV structure confirms that the N68 \HII region is kinematically associated with the N68a cloud, consistent with the expansion-driven morphology of bubble N68. 
Importantly, the observed bridges are located away from the influence of the \HII region (see also Figure \ref{fig:pv}(a)), indicating that their origin is not related to the bubble’s expansion. 
Instead, their kinematics and location are more consistent with a CCC interface.

\subsubsection{Complementary distribution} \label{subsec:comp}

\begin{figure}
    \centering
    \includegraphics[width=0.8\textwidth]{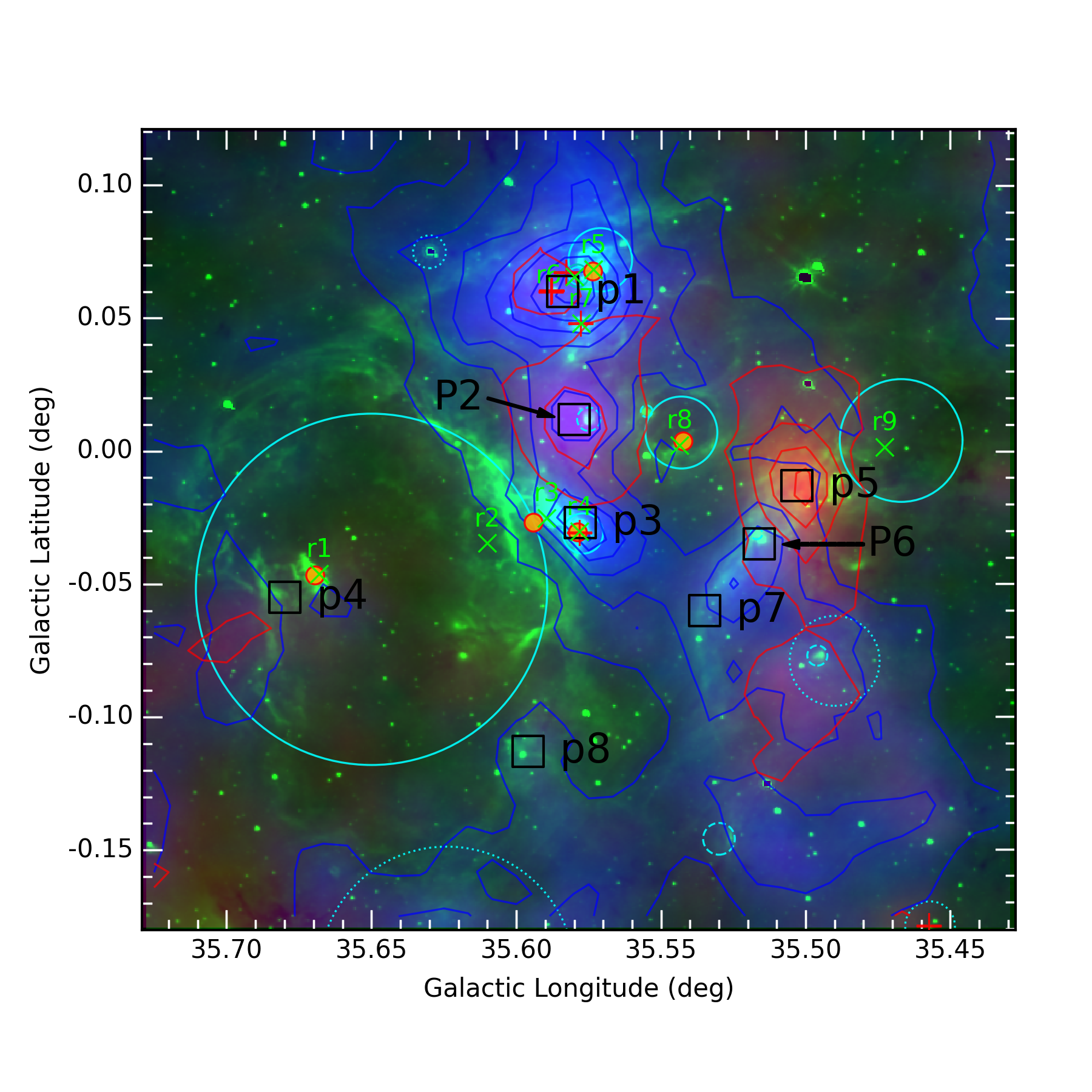}
    \caption{Complementary distribution of N68a and N68b. The RGB-composite map with its R, G and B colors encoded by N65b, $Spitzer-$IRAC 8.0 $\mu$m and N65a, respectively. The solid blue contours indicate the $^{13}$CO integrated-intensity map of N65a, while the dashed red contours indicate that of N65b. The contour levels in the figure are drawn from 9 (3$\sigma$) to 45 K km s$^{-1}$ in steps of 6 K km s$^{-1}$. }
    \label{fig:comp}
\end{figure}

A complementary spatial distribution between two colliding clouds provides additional morphological evidence for CCC \citep{Fukui2018b, Takahira2018}. 
Figure \ref{fig:comp} shows the spatial distributions of N68a (blue) and N68b (red), overlaid on the 8 $\mu$m emission (green). 
The two clouds exhibit a clear overlapping and complementary structure. 
N68a exhibits a more extended structure, encompassing clumps p1, p2, p3, p4, p6, p7, and p8, while N68b is more compact, containing only clumps p2, p5, and p6. 
The clouds overlap directly at the interface positions p2 and p6. 
Notably, clumps p1, p2, and p3 in N68a are arranged in a filamentary structure, while the N68b clump at p2 appears precisely inserted between p1 and p3, demonstrating a complementary distribution. 
A similar complementary pattern is seen at p6, which lies between clump p5 (N68b) and clump p7 (N68a). 
Although no clear massive star formation tracers (e.g., \HII regions or masers) are detected at p2 and p6, both positions coincide with strong 8 $\mu$m emission sources, indicating embedded low- or intermediate-mass star formation. 
These spatial and kinematic patterns are naturally explained by a CCC scenario in which N68a and N68b collide at p2 and p6, compressing the gas and triggering star formation at the interface.

\subsection{Induced star formation} \label{subsec:SF}

\begin{figure}
    \centering
    \includegraphics[width=0.9\textwidth]{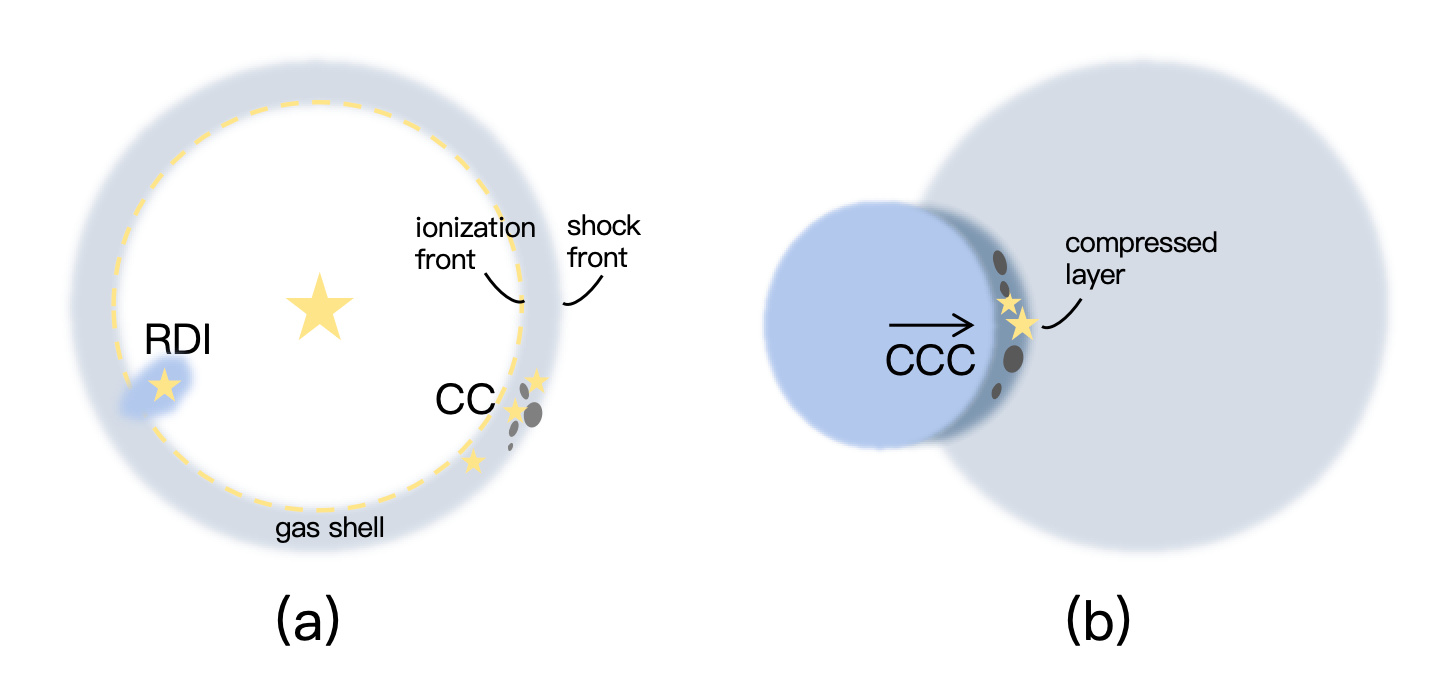}
    \caption{Schematics of induced star formation mechanisms in terms of CC, RDI and CCC, respectively. }
    \label{fig:schematic}
\end{figure}

Based on the results and discussions presented, we conclude that star formation in the N68 region is driven by multiple concurrent mechanisms. 
These include the collect-and-collapse (CC) process triggered by the expansion of an evolved \HII region, the radiation-driven implosion (RDI) acting on pre-existing dense cores at the edges of ionization fronts, and cloud-cloud collision (CCC) at the interface between two independent molecular clouds. 
The coexistence and interplay of these three mechanisms have collectively induced star formation across a wide range of masses within the R1 region of N68, from massive O/B-type stars to low- and intermediate-mass YSOs. 
We use a simple schematic (see Figure \ref{fig:schematic}) to illustrate these three trigger mechanisms. 
Figure \ref{fig:schematic} (a) shows the CC mechanism and the RDI mechanism caused by the expansion of the \HII region. 
Although both of these mechanisms are caused by irradiation of an external source of ionizing photons, their targets are different. 
The RDI process requires pre-existing dense clumps, although these clumps would inevitably collapse to form stars anyway, the ionization source plays a triggering role, accelerating this process \citep{Dale2007b}. 
The mechanism of CC is that the gas shell driven by the \HII region fragments to form stars \citep{Dale2007a}. 
The CCC mechanism is similar to the CC, which also involves a process of ``collect and collapse", with the driving force originating from the kinetic movement between molecular clouds rather than ionization expansion. 
The collision model was first proposed by \cite{Habe1992}, as shown in Figure \ref{fig:schematic} (b), the interface between the supersonic collision clouds was compressed to form a dense layer, which intern fragments and collapses to form stars.

Bubble N68 displays a well-defined shell-like morphology, with the brightest infrared emission concentrated along its eastern (Cloud 4 in Figure \ref{fig:CTG}) and western (Cloud 3) edge, respectively. 
Based on the characteristics of CC and RDI, the star formation in Cloud 3 is dominated by the CC mechanism, while in Cloud 4 it is dominated by the RDI mechanism. 
According to Tables \ref{tab:bubbles} and \ref{tab:clouds}, Cloud 3 contains a B0V star (r2), and 6 Class I and 16 Class II YSOs, with star formation efficiency (SFE) about 1.4\%. 
For comparison, Cloud 4 harbors a B0V star (r1), 1 MYSO, and 5 Class I and 2 Class II YSOs, with SFE about 2.1\%. 
From the results, it seems that these two mechanisms have triggered similar star formation. 
However, the stars formed by the RDI mechanism are relatively younger, indicating that, in the same ionization source-driven CC process and RDI process, the star formation triggered by RDI may occur later. 
Cloud 4 exhibits a prominent bright-rimmed cloud (BRC) structure with its bright ``head” orients toward the interior of the bubble, while the fainter, more diffuse ``tail" extends eastward, indicating the direction of ionization and shock propagation. 
This eastward gas motion is also evident in the PV slices L3 and L4 presented in Figure \ref{fig:spectra}. 
A distinct velocity gradient observed at p4 reflects a kinematic difference between the irradiated ``head" and the more quiescent ``tail" of the BRC, a signature consistent with compression by an ionization front.

The CCC mechanism occurs mainly in Cloud 1 and Cloud 2. 
The star formation in Cloud 1 is very diverse, including 2 O9.5V massive stars (r3 and r5), 4 B-type stars (r4 and r6-8), 4 MYSOs, and 16 Class I and 16 Class II YSOs. 
However, the number of massive stars formed by Cloud 2 is relatively small, including one B0.5V star (r9), and 16 Class I and 44 Class II YSOs. 
It is worth noting that not all of these stars could have been triggered by CCC, for example, r3 and r4, which are adjacent to r2, might have been affected by the CC process. 
Nevertheless, even if we assume that r3 and r4 were triggered by CCC, the calculated SFE of Cloud 1 and Cloud 2 were only 1\% and 0.8\%, respectively, which were much lower than that of the nearby molecular clouds (3-6.3\% by \citealt{Evans2009}). 
This seems to indicate that CCC has not significantly enhanced the star formation efficiency. 
However, considering that Cloud 1 has 6 O/B-type stars and 4 MYSO candidates, CCC seems to be more inclined to form massive stars, which is consistent with the results of \cite{Fukui2021}.

\cite{Enokiya2021} suggested that the number of O-type stars triggered by CCC is related to a threshold column density, with the empirical fitting formula as $log(N_{\rm O})=0.73\ log(N_{\rm H_2})-16.1$. 
Based on this, we estimate that the number of O-type stars that could potentially form in Cloud 1 is 2.3, which is comparable to the two O9.5V stars contained in Cloud 1. 
For Cloud 2, it was expected to form 1.6 O-type stars, but none were found. 
This might be related to the fact that the collision occurred very recently. 
As shown in Figures \ref{fig:comp} no obvious signs of massive star formation were found at the collision interfaces p2 and p6. 
However, it is notable that both locations have relatively bright 24 and 70 $\mu$m emissions (as shown in Figure \ref{fig:overlay}), and each is associated with a radio-quiet \HII region \citep{Anderson2014}. 
These characteristics suggest that the collision interface may be hosting ongoing massive star formation, possibly in an early, deeply embedded phase not yet detectable in radio continuum. 
These signatures collectively suggest that the collision between the N68a and N68b clouds is a relatively recent event. 
Due to the relatively small contact area (or resulting compressed cavity) between the two molecular clouds, it is difficult to estimate the collision timescale based on their kinematic displacement. 
We therefore adopt the characteristic age of Class I YSOs ($\sim$0.44 Myr; \citealt{Evans2009}) as an approximate upper limit for the collision timescale, while the lower limit could be considerably more recent. 
This short duration of the collision ($<0.44$ Myr) and the relatively small collision area (p2 and p6) suggest that the CCC is unable to affect the formation of the evolved N68 bubble (dynamical age $>2.9$ Myr).

\subsection{Large scale collisions} \label{subsec:pc}

\begin{figure}
    \centering
    \includegraphics[width=0.95\textwidth]{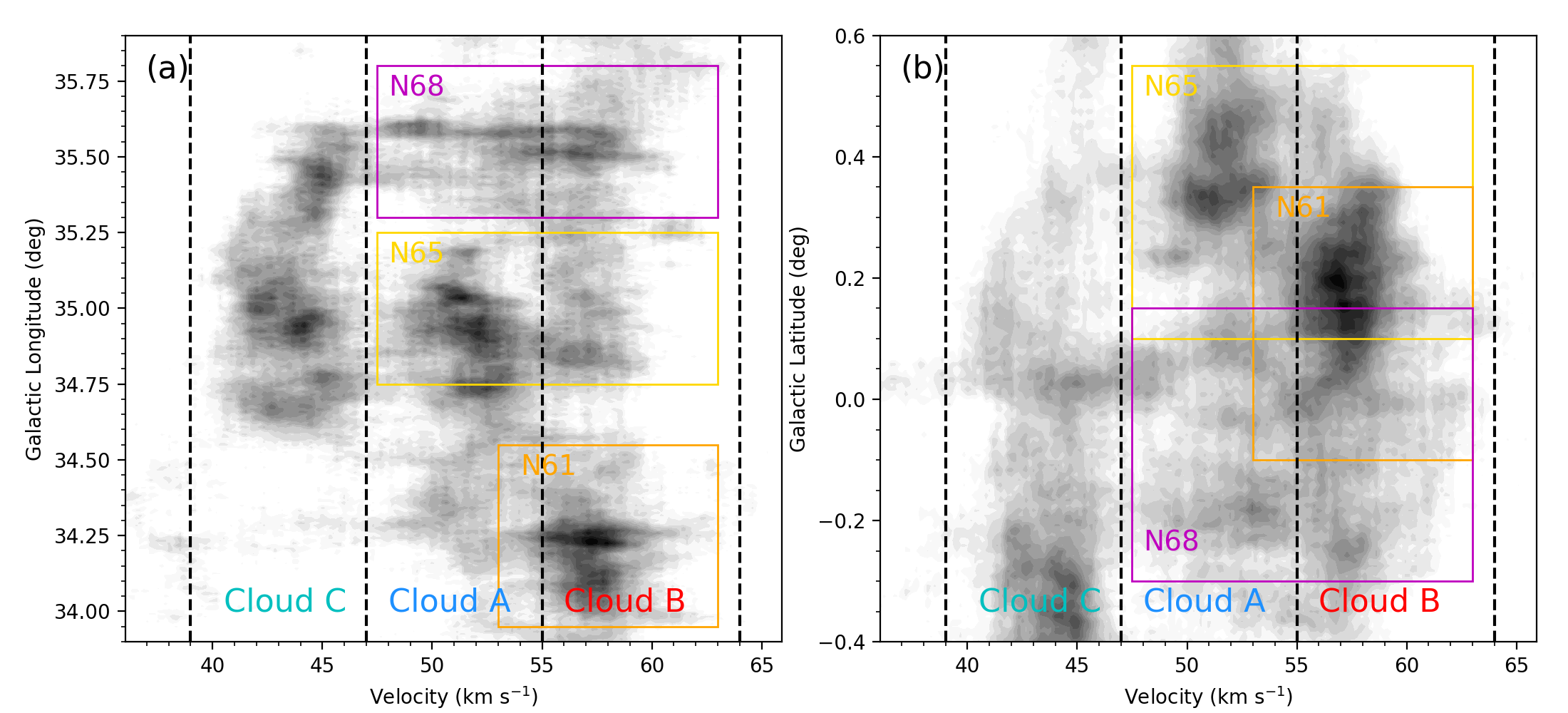}
    \includegraphics[width=1\textwidth]{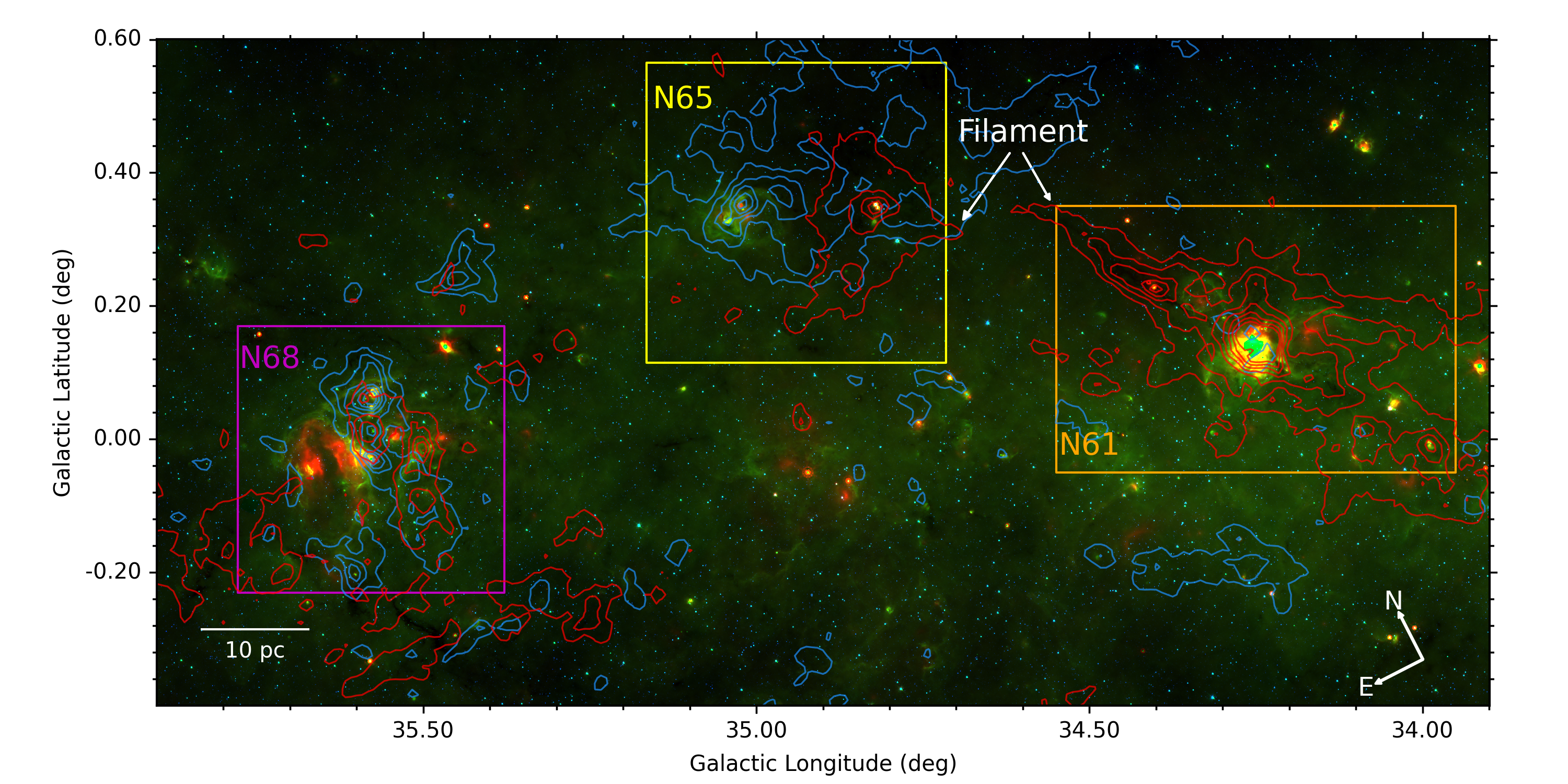}
    \caption{Overview of large scale gas distribution around the G35 complex. Panels (a) \& (b) are $^{13}$CO $l-$PV and $b-$PV diagrams, respectively. The vertical dashed line divided the gas into three individual components, namely Cloud A [47, 55] \kms, Cloud B [55, 64] \kms and Cloud C [39, 47] \kms. The magenta, yellow and orange boxes indicate the counterpart gas with bubbles N68, N65 and N61, respectively. Panel (c) is the $^{13}$CO gas distribution in terms of Cloud A (blue) and Cloud B (red), respectively. The three colored boxes are the regions of N68 (magenta), N65 (yellow) and N61 (orange), respectively. The background image is the RGB-composite image (R: 24 $\mu$m, G: 8.0 $\mu$m and B: 3.6 $\mu$m).  }
    \label{fig:large-scale}
\end{figure}

A large-scale view of bubble N68 reveals a number of notable features. 
Figure \ref{fig:large-scale}(c) shows the distribution of the molecular cloud surrounding bubble N68, which is also part of the G35 molecular cloud complex. 
This complex contains three main bubble structures—N68 (magenta), N65 (yellow), and N61 (orange)—all continuously connected in molecular emission. 
Position-velocity diagrams (Figures \ref{fig:large-scale}(a) and (b)) show that the molecular gas linked to these bubbles spans 47–64 \kms. 
Based on velocity coherence, the G35 complex can be decomposed into two large-scale sub-clouds: Cloud A (47–55 \kms) and Cloud B (55–64 \kms). 
Within these, N68 contains two colliding clouds, N68a (47–56 \kms) and N68b (56–64 \kms), which are part of Cloud A and Cloud B, respectively. 
Similarly, \cite{Chen2025} has reported that the nearby bubble N65 consists of colliding clouds N65a (47–55 \kms) and N65b (55–62 \kms). 
Given the velocity consistency, we infer that the collision between Cloud A and Cloud B occurred simultaneously in N68 and N65. 
For N61, although there is no obvious collision structure, its main component lies between 53–64 \kms and likely belongs to Cloud B. 
A filament connecting N61 and N65b further supports the spatial continuity of Cloud B. 
In short, Cloud A contains two molecular clouds of N68a and N65a, while Cloud B contains three molecular clouds of N68b, N65b and N61. 
According to the coordinates and distances of N68 (3.4 kpc), N65 (3.5 kpc) and N61 (3.7 kpc), we roughly estimated the extent of Cloud A and Cloud B to be 50 pc and 100 pc, respectively. 

Overall, we propose a large-scale, hierarchical CCC scenario in the G35 complex. 
The simultaneous collision signatures observed in both N68 and N65, which lie at different locations along the interface between Cloud A and Cloud B, indicate that this is not an isolated, localized event, but rather a coherent, large-scale interaction. 
This collision spans approximately 100 pc in extent, influencing multiple star-forming regions across the complex. 
Such a scenario suggests that CCC can operate over significantly larger spatial scales than typically considered, organizing and synchronizing star formation across pc-scale sub-regions. 
The continuity of Cloud B, evidenced by the filament connecting N61 and N65b, further supports the interpretation of a single, extended colliding structure. 
Therefore, the star formation activity in N68, N65, and likely N61, can be understood as a local manifestation of this overarching, large-scale collisional process, highlighting the importance of cloud-scale dynamics in regulating massive star formation across entire molecular complexes.

\cite{Fukui2018a} reported a similar 100 pc scale CCC (NGC 6334 and NGC 6357), which is continuous over $3^{\circ}$ along the galactic plane with a collision speed exceeding 12 \kms. 
The collisions of two large clouds triggered the formation of two mini-starbursts over 100 pc in the past few Myr. 
Compared to such a violent collision, our 100 pc collision in G35 seems much ``quieter", with a collision speed of 4-5 \kms, and few O stars were formed. 
The origin of such large-scale collisions is not very clear. 
We have noticed that the G35 molecular cloud is spatially close to the meeting point of the Scutum-Centaurus arm and the Galactic bar. 
The gas in the spiral arms tends to accumulate at this meeting point, leading to a starburst there (W43, \citealt{Zhang2014}). 
Therefore, the gravitational potential of the Galactic bar may play a crucial role in the large-scale collisions of G35.

\section{Conclusions} \label{sec:conclusion}

N68 is an active and complex massive star formation region, featuring numerous indicators of high-mass star formation, such as \HII regions, radio peaks, $MSX$ sources, 6.7 GHz maser sources and a BRC, as well as low-/intermediate-mass star formation, such as MIR (24 $\mu$m and 70 $\mu$m) emission sources and YSOs. 
Our comprehensive analysis indicates that the star formation process of N68 is the result of the combination of multiple star formation mechanisms, such as CC, RDI and CCC. 
The main conclusions of this work are summarized as follows: 

\begin{enumerate}
 \item Evidence for Cloud-Cloud Collision: Two distinct molecular clouds (N68a: 47–56 \kms and N68b: 56–64 \kms) are identified toward N68. Their complementary distribution and the broad velocity bridge at border clumps (p2 and p6) provide direct evidence for an ongoing CCC. 

 \item Young Stellar Populations: We identified ten O/B-type stars and 163 YSOs (45 Class I, 5 Flat-spectrum, 113 Class II), indicating active star formation across a wide mass range. 
 
 \item Multiple Star Formation Mechanisms: Star formation in N68 is triggered collectively by three mechanisms — CC, RDI, and CCC. These processes are spatially separated: a) CC in the bubble periphery (Cloud 3) formed a B0V star. b) RDI in a bright rim cloud (Cloud 4) produced a B0V star. c) CCC at the cloud interface (clumps p2 and p6) likely triggered two radio-quiet \HII regions. 

 \item Star Formation Efficiency: The CCC mechanism does not enhance the star formation efficiency, but is more likely to trigger the formation of massive stars. 

 \item Large-Scale CCC System: N68, together with neighboring bubbles N65 and N61, forms a $\sim 100$ pc scale CCC system in the G35 complex, suggesting that CCC operates over extended spatial scales to organize star formation across the region. 

\end{enumerate}

\begin{acknowledgments}
We acknowledge support from the National Key R\&D Program of China (2022YFA1603102) and the National Natural Science Foundation of China (NSFC grant No. 12473024). 
X.C. acknowledges the support of the Guangdong Province Universities and Colleges Pearl River Scholar Funded Scheme (2019). 
This work made use of data products from the Milky Way Imaging Scroll Painting (MWISP) project, which is sponsored by the National Key R\&D Program of China (Grants No. 2023YFA1608000 and 2017YFA0402701) and by the CAS Key Research Program of Frontier Sciences (Grant No. QYZDJ-SSW-SLH047). 
We also made use of the near-infrared data from the UKIRT Infrared Deep Sky Survey (UKIDSS), mid-infrared data from the $Spitzer$ Space Telescope (operated by the Jet Propulsion Laboratory, California Institute of Technology under NASA contract), and far-infrared $Herschel$ images at 70, 160, 250, 350, and 500 $\mu$m from the $Herschel$ Infrared Galactic Plane Survey (Hi-GAL). 
$Herschel$ is an ESA space observatory with science instruments provided by European-led Principal Investigator consortia and with important participation from NASA. 
We also used the 870 $\mu$m dust continuum data from the APEX Telescope Large Area Survey of the Galaxy (ATLASGAL). 
The 5 GHz (CORNISH) and 1.4 GHz (MAGPIS and NVSS) radio continuum data are extracted from the Very Large Array (VLA) radio interferometer surveys of the Galactic Plane, and the 6.7 GHz methanol maser data from the online tool $MaserDB$. 
\end{acknowledgments}

\bibliography{library}{}
\bibliographystyle{aasjournal}



\end{document}